\documentclass[twoside,11pt,leqno]{article}
\usepackage{jair, theapa, rawfonts}

\usepackage{tabularx}
\usepackage{multirow}
\usepackage{multicol}
\usepackage{graphicx}
\usepackage{color}
\usepackage{amssymb}
\usepackage{amsmath}
\usepackage{amsthm}
\usepackage{booktabs}
\usepackage{tikz}
\usetikzlibrary{positioning}
\usetikzlibrary{calc}
\usepackage{enumerate}
\usepackage{rotating}
\usepackage[linesnumbered,vlined,ruled]{algorithm2e}
\usepackage{etoolbox}
\usepackage{thm-restate}

\newcommand{\OMIT}[1]{} %

\newcommand{\card}[1]{|#1|}

\newcommand{\p}{\mbox{\rm P}}

\newcommand{\np}{\mbox{\rm NP}}

\newcommand{\daa}{\textit{AA}\xspace}
\newcommand{\dlcd}{\textit{LCD}\xspace}
\newcommand{\dcd}{\textit{CD}\xspace}
\newcommand{\dvd}{\textit{VD}\xspace}
\newcommand{\dgs}{\textit{GS}\xspace}
\newcommand{\dls}{\textit{LS}\xspace}
\newcommand{\dcp}{\textit{CP}\xspace}
\newcommand{\dc}{\textit{CL}\xspace}
\newcommand{\dw}{\textit{WI}\xspace}
\newcommand{\vote}[1]{\ifstrempty{#1}{\succ}{\succ_{#1}}}

  \newtheorem{theorem}{Theorem}[section]

  \newtheorem{lemma}[theorem]{Lemma}
  \newtheorem{observation}[theorem]{Observation}
  \newtheorem{remark}[theorem]{Remark}
  
  \newtheorem{proposition}[theorem]{Proposition}
  \newtheorem{definition}[theorem]{Definition}
  \newtheorem{example}{Example}

\newcommand{\EP}[3]{
\begin{center}
{\small 
\begin{tabularx}{0.98\columnwidth}{ll}
\toprule
\multicolumn{2}{c}{\textsc{#1}} \\
\midrule
{\bf Given:}   & \parbox[t]{0.8\columnwidth}{#2\vspace*{1mm}}  \\
{\bf Question:}& \parbox[t]{0.8\columnwidth}{#3\vspace*{.5mm}} \\ 
\bottomrule
\end{tabularx}
}
\end{center}
}

\newcounter{continueexcnt}

\newcounter{counterexno}
\newcommand{\counterex}[2]{\refstepcounter{counterexno}\textit{Counterexample \thecounterexno\ (#2):}\ \ \label{#1}}

\newcommand{\circlesucc}{ 
  \mathbin{
    \mathchoice
      {\tikzcirclesucc{\displaystyle}}
      {\tikzcirclesucc{\textstyle}}
      {\tikzcirclesucc{\scriptstyle}}
      {\tikzcirclesucc{\scriptscriptstyle}}
  } 
}

\newcommand\tikzcirclesucc[1]{%
  \begin{tikzpicture}[baseline=(X.base), inner sep=-0.4, outer sep=0]
    \node[draw,circle] (X)  {$#1\,\succ$};
  \end{tikzpicture}%
}

\jairheading{58}{2017}{297-337}{05/16}{02/17} 
\ShortHeadings{Computational Aspects of Nearly Single-Peaked Electorates}{Erd\'{e}lyi, Lackner \& Pfandler}
\firstpageno{297}

\begin{document}
\title{Computational Aspects of Nearly Single-Peaked Electorates}

\author{\name G\'{a}bor Erd\'{e}lyi \email erdelyi@wiwi.uni-siegen.de \\
       \addr School of Economic Disciplines\\
       University of Siegen\\
       Siegen, Germany
       \AND
       \name Martin Lackner \email martin.lackner@cs.ox.ac.uk \\
       \addr Department of Computer Science\\
       University of Oxford \\
       Oxford, United Kingdom
       \AND
       \name Andreas Pfandler \email pfandler@dbai.tuwien.ac.at \\
       \addr Institute of Information Systems\\
       TU Wien \\
       Vienna, Austria%
}

\maketitle

\begin{abstract}
Manipulation, bribery, and control are well-studied ways of changing the 
outcome of an election. Many voting rules are, in the general case, 
computationally resistant to some of these manipulative actions. However 
when restricted to single-peaked electorates,
these rules suddenly become easy to manipulate. Recently, Faliszewski, 
Hemaspaandra, and Hemaspaandra
studied the computational complexity of strategic behavior in nearly single-peaked 
electorates. These are electorates that are not single-peaked but 
close to it according to some distance measure.

In this paper we introduce several new distance measures regarding 
single-peakedness. We prove that determining whether a given profile is 
nearly single-peaked is \np-complete in many cases.
For one case we present a polynomial-time algorithm.
In case the single-peaked axis is given, we show that determining the distance is always possible in polynomial time.
Furthermore, we  explore the relations between the new notions introduced in this paper and existing notions from the literature.
\end{abstract}

\section{Introduction}
\label{sec:introduction}

Voting is a ubiquitous method for preference aggregation and collective decision-making.
It has applications in many settings ranging from politics to artificial intelligence and further topics in computer science \cite<see, e.g.,>{eph-ros:j:multiagent-planning,gho-mun-her-sen:c:voting-for-movies,dwo-kum-nao-siv:c:rank-aggregation}. In the presence of huge data volumes, the computational properties of voting rules gain great importance. In particular, it is desirable to be able to quickly determine the winner(s) of an election.
On the other hand it should be computationally hard to find strategies for dishonest behavior.

\citeA{btt89} were the first to study the computational aspects of strategic behavior in elections.
They defined and studied manipulation in voting, i.e., a group of voters casts their votes insincerely in order to reach a desired outcome.
Another type of manipulative behavior is control, where an external agent makes structural changes to the election such as adding/deleting/partitioning either candidates or voters in order to reach a desired outcome.
Control has been studied first also by \citeA{bartholdi-control}.
There is also bribery, where an external agent changes some votes in order to influence the outcome of the election \cite<see, e.g.,>{fal-hem-hem:j:bribery}. For an overview and many natural examples of bribery, control, and manipulation we refer to the literature~\cite{bau-erd-hem-hem-rot:b:computational-apects-of-approval-voting,fal-hem-hem:j:complexity-protect,fal-pro:j:manipluation,bra-con-end:b:comsoc,rot:b:comsoc,handbook:comsoc}.

Traditionally, the complexity of such ``attacks'' is studied under the assumption that, in each election, any admissible vote can occur. However, there are many elections where the diversity of the votes is limited in the sense that there are admissible votes nobody would ever cast. 
One of the best known examples is the \emph{single-peaked} domain, introduced by \citeA{black}.
It is based on the assumption that the votes are polarized along some linear axis and voters prefer candidates closer to their ideal candidate on this axis over candidates farther away.
The study of the computational aspects of elections with single-peaked preferences was initiated by \citeA{wal:c:uncertainty}, followed by fundamental contributions by \citeA{fal-hem-hem-rot:j:shield} and \citeA{bra-bri-hem-hem:j:single-peaked}.
The general conclusion of these papers is that many voting problems which are $\np$-hard in the general case turn out to be easy for single-peaked societies.

A recent line of research initiated by \citeA{con:j:single-peaked} suggests that many elections are not perfectly single-peaked but are \emph{close} to it with respect to some measure.
In the work of \citeA{fal-hem-hem:j:nearly-single-peaked} various notions of nearly single-peaked elections were introduced and it was shown that the complexity of manipulative actions jumps back to \np-hardness in many cases.

This paper is the first to systematically study notions of distances for nearly single-peaked electorates.
Our main contributions are:
\begin{itemize}
\item We introduce three new notions of nearly single-peakedness. In addition, we study six notions that already have been defined or suggested in the literature.
\item We explore connections between both existing and new notions by providing inequalities.
These allow one to compare these notions and better understand their relationship. We also briefly discuss to which degree axiomatic properties of single-peaked preferences transfer to nearly single-peaked preferences.
\item We analyze the computational complexity of computing the distance of arbitrary preference profiles to single-peakedness.
In most cases we show \np-completeness.
For the $k$-candidate deletion distance, we present a polynomial-time algorithm.
\item Finally, we analyze the computational complexity of the nearly single-peaked evaluation problem, where the task is to compute the distance for a given axis.
\end{itemize}

\subsection{Related Work}
The main computational problem studied in this work is recognizing nearly single-peaked preferences. For single-peaked preferences, the question of whether a given profile is single-peaked has received considerable attention. 
\citeA{bar-tri:j:psychological} were the first to prove that it requires only polynomial time to detect single-peaked preferences; their algorithm based on the consecutive ones problem requires $\mathcal{O}(m^2n)$ time for profiles with $n$ voters and $m$ candidates. This runtime was improved to $\mathcal{O}(mn+m^2)$ by \citeA{doignon1994polynomial} and finally to $\mathcal{O}(mn)$ by \citeA{esc-lan-oez:c:single-peak}.
Our work, in contrast to these results, shows that computing the distance to single-peaked preferences is often computationally hard.
The impact of single-peaked preferences on computational problems in social choice is generally well understood. Let us mention
the work of \citeA{fal-hem-hem-rot:j:shield} and \citeA{bra-bri-hem-hem:j:single-peaked}, in which the complexity of winner problems and of strategic behavior (e.g., manipulation and control) in electorates with single-peaked preferences is investigated. These papers do not consider nearly single-peaked preferences, but mention them as future work.

In the context of nearly single-peaked preferences the most relevant paper is by \citeA{fal-hem-hem:j:nearly-single-peaked}. They introduce several notions of nearly single-peakedness and analyze the complexity of bribery, control, and manipulation in nearly single-peaked elections.
Their work on manipulation and bribery has recently been extended to other voting rules \cite{DBLP:conf/aaai/MenonL16} and to other notions of distance \cite{adt/ErdelyiLP15-manipulation}.
These papers deal with applications of nearly single-peaked electorates and assume that the underlying axis is part of the input.
Our paper, in contrast, studies the complexity of computing an axis that minimizes the distance of a preference profile to single-peakedness.
The work of \citeA{bre-che-woe-profiles-nearby} has a similar objective but is somewhat orthogonal to our paper. They study only two distance measures (Voter Deletion and Candidate Deletion) but for a variety of domain restrictions (e.g., also the single-crossing and group-separable domains).
Their work is based on a combinatorial characterization of domain restrictions by \citeA{bal-hae:j:characterization-single-peaked} and \citeA{bre-che-woe:j:single-crossing}.
Recent work by \citeA{aaai/ElkindL14-approx} presents approximation and fixed-parameter tractable (fpt) algorithms for computing such distances.
\citeA{sui2013multi} propose heuristics to compute distances to single-peakedness and its two-dimensional analogue, 2D single-peakedness; they also perform experiments on real-world data sets.
Two further distance measures have been studied:
Single-peaked width by \citeA{cor-gal-spa:c:single-peaked-width,DBLP:conf/ijcai/CornazGS13} and 
the decloning measure by \citeA{elk-fal-sli:c:clone-structures}.
The complexity of manipulation and control in profiles of bounded single-peaked width has been studied by \citeA{yang2014controlling} and \citeA{yang2015manipulation}.

Single-peakedness on trees \cite{demange1982single} and single-peakedness on circles \cite{aaai/PetersL17-spoc} are also generalizations of classical single-peakedness but are not directly related to (distance based) nearly single-peakedness. Both concepts have proven to be algorithmically useful in the context of multiwinner elections \cite{DBLP:conf/ijcai/YuCE13,peters2016nicetrees,aaai/PetersL17-spoc}.
Multiwinner elections have also been studied for single-peaked \cite{BSU13a} and single-crossing elections~\cite{DBLP:journals/tcs/SkowronYFE15}.

Another line of research are domain restrictions in incomplete preferences (partial orders); incomplete single-peaked profiles \cite{aaai/Lackner14-incompletesp} and incomplete single-crossing profiles \cite{aaai/ElkindFLO15-incompletesc} have been considered. Domain restrictions have also been studied in dichotomous preferences~\cite{aaai/ElkindL15-dichpref} and notions of single-peakedness for preferences with ties, i.e., for weak orders \cite{fitzsimmons2016modeling}.
Finally, we remark that single-peaked preferences have also been considered in the context of preference elicitation~\cite{con:j:single-peaked} and in the context of possible and necessary winners under uncertainty regarding the votes~\cite{wal:c:uncertainty}. The mathematical likelihood of single-peaked preferences has been analyzed by \citeA{arxiv/BrunerL-likelihoodSP}.

\subsection{Organization}

This paper is organized as follows. In Section~\ref{sec:prelims}, we recall some notions from voting theory and define single-peaked profiles.
In Section~\ref{sec:problem-statement}, we introduce the decision problems we are investigating in our paper.
Section~\ref{sec:results-basic} presents basic results regarding single-peaked profiles. Our results on the relations between the different notions of nearly single-peakedness are presented in Section~\ref{sec:relations}.
The results on the complexity of nearly single-peaked consistency and evaluation can be found in Section~\ref{sec:complexity-cons}.
Finally, Section~\ref{sec:conclusions} provides some conclusions and directions for future work.

\section{Preliminaries}
\label{sec:prelims}

Let $C$
be a finite set of \emph{candidates} %
and let $\vote{}$ be a total order on $C$.
Let $\mathcal{P}=(\vote{1}, \ldots , \vote{n})$ be a \emph{preference profile}, i.e., a list of total orders on the candidate set $C$. 
An \emph{election} is defined as a pair $E=(C,\mathcal{P})$, where $C$ is the set of candidates and $\mathcal{P}$ a preference profile on $C$.
We say that $\vote{i}$ is the \emph{vote} of voter $i$. For simplicity we write $\vote{i}: c_1c_2\ldots c_m$ instead of $c_1 \succ_i c_2 \succ_i \cdots \succ_i c_m$. For a vote $\vote{i}: c_1c_2\ldots c_m$ let the vote $\overline{\vote{i}}: c_m c_{m-1}\ldots c_1$ denote the \emph{reverse} vote of $\vote{i}$.
For two preference profiles on the same set of candidates $\mathcal{P}=(\vote{1}, \ldots , \vote{n})$ and  $\mathcal{L}=(\vote{n+1}, \ldots , \vote{s})$, let $(\mathcal{P}, \mathcal{L})=(\vote{1}, \ldots , \vote{s})$ define the \emph{union} of the two preference profiles. 
In our constructions, we sometimes insert a subset of the candidates $B\subseteq C$ into a vote, where we assume some arbitrary, fixed order of the candidates in $B$ (e.g., $\succ_i: c_1 B c_3$ means that $c_1$ is the top-ranked candidate of voter $i$ and $c_3$ is the last-ranked candidate, whereas all $b\in B$ are ranked between $c_1$ and $c_3$).

\begin{definition}\label{def:single-peaked}
Let an \emph{axis} $A$ be a total order on $C$ denoted by $>$.
Furthermore, let $\vote{}$ be a vote with top-ranked candidate $c$.
The vote $\vote{}$ is \emph{single-peaked with respect to $A$} if for any $x,y\in C$, 
if $x>y>c$ or $c>y>x$ then $c\succ y\succ x$ has to hold.

A preference profile $\mathcal{P}$ is said to be \emph{single-peaked with respect to an axis $A$} if and only if each vote is single-peaked with respect to $A$. 
A preference profile $\mathcal{P}$ is said to be \emph{single-peaked consistent} if there exists an axis $A$ such that $\mathcal{P}$ is single-peaked with respect to $A$. 
\end{definition}

Let $C'\subseteq C$.
By $\mathcal{P}[C']$ we denote the profile $\mathcal{P}$ restricted to the candidates in $C'$.
Analogously if $A$ is an axis on $C$, we denote by $A[C']$ the axis $A$ restricted to candidates~in~$C'$.

Escoffier, Lang, and \"{O}zt\"{u}rk \citeyear{esc-lan-oez:c:single-peak} present an algorithm that decides whether a given preference profile is single-peaked consistent in time $\mathcal{O}(m\cdot n)$. Their algorithm improves upon the runtime of the original algorithm by \citeA{bar-tri:j:psychological}. The corresponding decision problem is defined as follows.

\EP{Single-Peaked Consistency}
{An election $E=(C,\mathcal{P})$.}
{Is $\mathcal{P}$ single-peaked consistent?}

If an axis is given additionally in the input, we have the evaluation problem. Since \textsc{Single-Peaked Consistency} can be solved in $\mathcal{O}(m\cdot n)$ time, so can \textsc{Single-Peaked Evaluation}.

\EP{Single-Peaked Evaluation}
{An election $E=(C,\mathcal{P})$ and an axis $A$.}
{Is $\mathcal{P}$ single-peaked with respect to $A$?}

\section{Nearly Single-Peaked Preferences}
 \label{sec:problem-statement}

In real-world settings one has to expect a certain amount of ``noise'' in preference data.
The single-peakedness property is very fragile and thus susceptible to such noise.
The following example illustrates the \emph{fragility of single-peakedness}:
Consider the single-peaked election consisting of two kinds of votes: 
$a b c d$ and $d c b a$.
Assume that both votes have been cast by a large number of voters.
This election is single-peaked only with respect to the axis $a > b > c > d$ and its reverse.
Adding a single vote $a b d c$ destroys the single-peakedness property although this vote is almost identical to the first kind of votes.

In this section we formally define different notions of nearly single-peakedness.
All these notions define a distance measure\footnote{We remark that we use the words ``distance'' and ``distance measure'' with their informal meaning and not in the mathematical sense of a metric.} to single-peaked profiles.
We will now describe them with help of a running example and provide first (trivial) upper bounds on these distances.

The following notions of nearly single-peakedness are defined for profiles.
The same definitions also hold for elections and thus we do not strictly distinguish between elections and profiles.
Throughout the following definitions let $E=(C,\mathcal{P})$ be an election and $k$ be a positive integer.

\subsubsection*{\emph{$k$-Voter Deletion} (\dvd)}

The first formal definition of 
nearly single-peaked societies was 
given by \citeA{fal-hem-hem:j:nearly-single-peaked}, however the idea of removing  voters that are not single-peaked dates back to \citeA{con:j:single-peaked}.
Consider a preference profile $\mathcal{P}$ for which most voters are single-peaked with respect to some axis $A$.
The voters that are not single-peaked with respect to $A$ are referred to as \emph{mavericks} by \citeA{fal-hem-hem:j:nearly-single-peaked}.
The number of mavericks, i.e., the number of voters that have to be deleted, defines a natural distance measure to single-peakedness.
If an axis can be found for a large subset of the voters, this is still a fundamental observation about the structure of the preference profile.

\begin{definition}[\citeR{fal-hem-hem:j:nearly-single-peaked}]%
A profile $\mathcal{P}$ is \emph{$k$-voter deletion single-peaked with respect to an axis $A$}
if by removing at most $k$ votes from $\mathcal{P}$ one can obtain a preference profile $\mathcal{P}'$ that is single-peaked with respect to $A$.
Furthermore, $\mathcal{P}$ is \emph{$k$-voter deletion single-peaked consistent}
if there exists an axis $A$ such that $\mathcal{P}$ is \emph{$k$-voter deletion single-peaked with respect to $A$}.
Let $\dvd(\mathcal{P})$ denote the smallest $k$ such that $\mathcal{P}$ is $k$-voter deletion single-peaked consistent.
\end{definition}

Note that $\dvd(\mathcal{P})\leq n-1$ always holds.
We remark that $k$-voter deletion single-peaked is also referred to as $k$-maverick-SP~\cite{fal-hem-hem:j:nearly-single-peaked} and as $k$-maverick single-peaked consistent~\cite{erd-lac-pfa:c:single-peaked-aaai}.

\begin{example}\label{ex:notions}
  Consider an election with $C=\{a,b,c,d,e\}$ and $\mathcal{P}=\{\vote{1}, \vote{2}, \ldots, \vote{202}\}$. 
  We define $\vote{1}: a b c e d$, $\vote{2}: e d c a b$, the votes $\vote{3}\ldots\vote{102}: a b c d e$, and the remaining votes $\vote{103}\ldots\vote{202}: e d c b a$.
  Notice that any preference profile containing $a b c d e$ and $e d c b a$ may only be single-peaked consistent with respect to the axis $a>b>c>d>e$ and its reverse.
  Since $\vote{1}$ and $\vote{2}$ are not single-peaked with respect to this axis, $\mathcal{P}$ is not single-peaked.
  Deleting $\vote{1}$ and $\vote{2}$ yields single-peaked consistency and thus we have $\dvd(\mathcal{P})=2$.
\end{example}

\subsubsection*{\emph{$k$-Candidate Deletion} (\dcd)}
As suggested by \citeA{esc-lan-oez:c:single-peak}, let us consider deleting candidates to obtain a single-peaked profile.
This distance measure can be particularly useful if there are candidates that do not have ``a~correct place'' on any axis.
Examples could be candidates that are not well-known (e.g., a new political party) or candidates that prioritize other topics than most candidates and thereby are judged by voters according to different criteria.
The votes restricted to the remaining candidates might still have a clear and significant structure, in particular they might be single-peaked consistent.

\begin{definition}
A profile $\mathcal{P}$ is \emph{$k$-candidate deletion single-peaked with respect to an axis $A$}
if there exists a set $C'\subseteq C$ obtained by removing at most $k$ candidates from $C$ such that $\mathcal{P}[C']$ is single-peaked with respect to $A$. 
Furthermore, $\mathcal{P}$ is \emph{$k$-candidate deletion single-peaked consistent}
if there exists an axis $A$ such that $\mathcal{P}$ is \emph{$k$-candidate deletion single-peaked with respect to $A$}.
Let $\dcd(\mathcal{P})$ denote the smallest $k$ such that $\mathcal{P}$ is $k$-candidate deletion single-peaked consistent.
\end{definition} 

Note that $\dcd(\mathcal{P})\leq m-2$ always holds.

{\addtocounter{example}{-1}
\stepcounter{continueexcnt}%
\begin{example}[continued]
  Consider the preference profile $\mathcal{P}$ as defined above.
  Observe that for $C'=\{b,c,d\}$, $\mathcal{P}[C']$ is single-peaked consistent.
  Deleting a single candidate does not yield single-peaked consistency and thus $\dcd(\mathcal{P})=2$.
\end{example}}

\subsubsection*{\emph{$k$-Local Candidate Deletion} (\dlcd)}
Personal friendships or hatreds between voters and candidates could move candidates up or down in a vote.
These personal relationships cannot be reflected in a global axis; this is an obstacle to single-peakedness already discussed by \citeA{con:j:single-peaked}.
To eliminate the influence of personal relationships to some candidates we define a local version of the previous notion.
This notion can also deal with the possibility that the least favorite candidates might be ranked without special consideration or even randomly.

We first have to define partial domains and partial profiles. 

\begin{definition}
Let $C$ be a set of candidates and $A$ an axis on $C$. A vote $\vote{}$ on a candidate set $C'\subset C$ is called a partial vote.
It is said to be \emph{single-peaked with respect to $A$} if it is single-peaked with respect to $A[C']$. %
A partial preference profile consists of partial votes.
It is called \emph{single-peaked consistent} if there exists an axis $A$ such that its partial votes are single-peaked with respect to $A$.
\end{definition}

\begin{definition}
A profile $\mathcal{P}$ is \emph{$k$-local candidate deletion single-peaked with respect to an axis $A$}
if by removing at most $k$ candidates from each vote in $\mathcal{P}$ we obtain a partial preference profile $\mathcal{P}'$ that is single-peaked with respect to $A$.
Furthermore, $\mathcal{P}$ is \emph{$k$-local candidate deletion single-peaked consistent}
if there exists an axis $A$ such that $\mathcal{P}$ is \emph{$k$-local candidate deletion single-peaked with respect to $A$}.
Let $\dlcd(\mathcal{P})$ denote the smallest $k$ such that $\mathcal{P}$ is $k$-local candidate deletion single-peaked consistent.
\end{definition}

Note that $\dlcd(\mathcal{P})\leq m-2$ always holds.

{\addtocounter{example}{-1}
\stepcounter{continueexcnt}%
\begin{example}[continued]
  Note that it is sufficient to remove candidate $a$ from vote $\vote{1}$ and candidate $e$ from vote $\vote{2}$ to obtain single-peaked consistency.
  Consequently, $\dlcd(\mathcal{P})=1$.
\end{example}}

\subsubsection*{\emph{$k$-Additional Axes} (\daa)}
Another suggestion by \citeA{esc-lan-oez:c:single-peak} was to consider the minimum number of axes such that each vote is single-peaked with respect to at least one of these axes.
Note that this corresponds to partitioning the voters in such a way that each group of voters is single-peaked.
Additional Axes is particularly useful if each candidate represents opinions on several issues (as it is the case in political elections).
A voter's ranking of the candidates would then depend on which issue is considered most important by the voter and consequently each issue might give rise to its own corresponding axis.

\begin{definition}
\label{def:aa}
A profile $\mathcal{P}$ is \emph{$k$-additional axes single-peaked with respect to axes $A_1, \ldots,$ $A_{k+1}$}
if there exists a partition $\mathcal{P}_1, \ldots, \mathcal{P}_{k+1}$ of $\mathcal{P}$ such that the subprofile $\mathcal{P}_1$ is single-peaked consistent with respect to $A_1$, $\mathcal{P}_{2}$ is single-peaked with respect to $A_2$, etc.
Furthermore, $\mathcal{P}$ is \emph{$k$-additional axes single-peaked consistent}
if there exist $k+1$ axes $A_1, \ldots, A_{k+1}$ such that $\mathcal{P}$ is \emph{$k$-additional axes single-peaked with respect to $A_1, \ldots, A_{k+1}$}.
Let $\daa(\mathcal{P})$ denote the smallest $k$ such that $\mathcal{P}$ is $k$-additional axes single-peaked consistent.
\end{definition}

Note that $\daa(\mathcal{P})< \min\left(n,\frac{m!}{2}\right)$ always holds. This is because the number of distinct votes is bounded by $\frac{m!}{2}$, since at most $m!$ distinct votes exist and each vote and its reverse are single-peaked with respect to the same axis.

{\addtocounter{example}{-1}
\stepcounter{continueexcnt}%
\begin{example}[continued]
  Notice that $\vote{1}$ and $\vote{2}$ are single-peaked consistent with respect to axis $b>a>c>e>d$.
  The remaining votes are consistent with respect to $a>b>c>d>e$.
  Thus, one additional axis is sufficient and hence $\daa(\mathcal{P})=1$.
\end{example}}

\subsubsection*{\emph{$k$-Global Swaps} (\dgs)}
There is a second method of dealing with candidates that are ``not placed correctly'' according to an axis $A$. Instead of deleting them from either the candidate set $C$ or from a vote, we could try to move them to the correct position. We do this by performing a sequence of swaps of consecutive candidates.
We remark that the minimum number of swaps required to change one vote to another is the \emph{Kendall tau distance}~\cite{kendall} of these two votes.
For example, to get from vote $abcd$ to vote $adbc$, we first have to swap candidates $c$ and $d$, and then we have to swap $b$ and $d$.
The Global Swaps distance counts the number of swaps globally, i.e., it considers the total number of swaps required; this is in contrast to the Local Swaps distance.
Since swaps change a vote only in a subtle way, $k$-global swaps can be considered a less obtrusive notion than $k$-(local) candidate deletion.

\begin{definition}
A profile $\mathcal{P}$ is \emph{$k$-global swaps single-peaked with respect to an axis $A$}
if $\mathcal{P}$ can be made single-peaked with respect to $A$ by performing at most $k$ swaps of consecutive candidates in the profile.
Furthermore, we say that the profile $\mathcal{P}$ is \emph{$k$-global swaps single-peaked consistent}
if there exists an axis $A$ such that $\mathcal{P}$ is \emph{$k$-global swaps single-peaked with respect to $A$}.
Let $\dgs(\mathcal{P})$ denote the smallest $k$ such that $\mathcal{P}$ is $k$-global swaps single-peaked consistent.
\end{definition} 
Note that these swaps can be performed wherever we want -- we can have $k$ swaps in only one vote, or one swap each in $k$ votes.
Since rearranging a total order to obtain any other total order requires at most $\binom{m}{2}$ swaps, we know that $\dgs(\mathcal{P})\leq {\binom{m}{2}} \cdot n$. 

{\addtocounter{example}{-1}
\stepcounter{continueexcnt}%
\begin{example}[continued]
  It is possible to make $\mathcal{P}$ single-peaked consistent by swapping $d$ and $e$ in vote $\vote{1}$ and swapping $a$ and $b$ in vote $\vote{2}$.
  This gives $\dgs(\mathcal{P})=2$.
\end{example}}

\subsubsection*{\emph{$k$-Local Swaps} (\dls)}
We can also consider a ``local'', per-vote budget for swaps, i.e., we allow up to $k$ swaps per vote.
This distance measure has been introduced by \citeA{fal-hem-hem:j:nearly-single-peaked} as Dodgson$_k$.

\begin{definition}
A profile $\mathcal{P}$ is \emph{$k$-local swaps single-peaked with respect to an axis $A$}
if $\mathcal{P}$ can be made single-peaked with respect to $A$ by performing at most $k$ swaps per vote. 
Furthermore, $\mathcal{P}$ is \emph{$k$-local swaps single-peaked consistent}
if there exists an axis $A$ such that $\mathcal{P}$ is \emph{$k$-local swaps single-peaked with respect to $A$}. 
Let $\dls(\mathcal{P})$ denote the smallest $k$ such that $\mathcal{P}$ is $k$-local swaps single-peaked consistent.
\end{definition}

Note that $\dls(\mathcal{P})\leq \binom{m}{2}$ always holds. 

{\addtocounter{example}{-1}
\stepcounter{continueexcnt}%
\begin{example}[continued]
  Since only one swap is required in $\vote{1}$ and $\vote{2}$ each, we have $\dls(\mathcal{P})=1$.
\end{example}}

\citeA{fal-hem-hem:j:nearly-single-peaked} also introduce the PerceptionFlip$_k$ distance. An election $E=(C,\mathcal{P})$ is \emph{$k$-perception flip single-peaked with respect to an axis $A$}
if for every vote $V$ in $\mathcal{P}$, the axis $A$ can be transformed to an axis $A'$ by at most $k$ swaps of consecutive candidates so that $V$ is single-peaked with respect to $A'$. We show in the following lemma that PerceptionFlip$_k$ and $k$-Local Swaps are identical. In other words, we show that swapping consecutive candidates in the vote or in the axis has the same ``power''.

\begin{lemma}
Let $E=(C,\mathcal{P})$ be an election and $A$ an axis on $C$. The profile $\mathcal{P}$ is $k$-local swaps single-peaked with respect to $A$ if and only if $\mathcal{P}$ is $k$-perception flip single-peaked with respect to $A$.
\end{lemma}

\begin{proof}
Given two total orders on the candidate set $C$, we define a permutation $p(T_1, T_2)$ from $\{1,\ldots,m\}$ to $\{1,\ldots,m\}$ as follows:
$i$ maps to $j$ if the $i$-th largest element in $T_1$ equals the $j$-th largest element in $T_2$.
For $T_1:bac$ and $T_2:cab$ we have $p(T_1, T_2)=321$, where $321$ is a short form for the permutation $\{1\mapsto 3, 2\mapsto 2, 3\mapsto 1\}$.
Note that given a vote $V$ and an axis $A$, the vote $V$ is single-peaked with respect to $A$ if and only if $p(A,V)$ consists of a decreasing sequence followed by an increasing sequence. (The top-ranked candidate in $V$ corresponds to $1$ in the permutation, hence $V$ being single-peaked with respect to $A$ corresponds to $p(A,V)$ being a V-shaped sequence.)
Performing swaps in either $V$ or $A$ implies that the permutation $p(A,V)$ is permuted.
Swapping the $j$-th largest and the $(j+1)$-th largest element in $V$ implies that $j$ and $j+1$ are exchanged in $p(A,V)$.
Analogously, swapping the $i$-th and $(i+1)$-th largest element on $A$ implies that in $p(A,V)$ the elements in position $i$ and $i+1$ are exchanged.
If we view a sequence of swaps as a permutation $\sigma$, then the number of swaps is equal to the number of inversions in $\sigma$, i.e., the number of pairs $i<j$ with $\sigma(i)>\sigma(j)$.
In the case of swaps in the vote, the permutation $\sigma$ is directly applied to $p(A,V)$; in the case of swaps in the axis the inverse $\sigma^{-1}$ is applied to $p(A,V)$.
Consequently, a vote $V$ can be made single-peaked with respect to $A$ by at most $k$ swaps if and only if there exists a permutation $\sigma$ with at most $k$ inversions such that $\sigma$ applied to $p(A,V)$ consists of a decreasing sequence followed by an increasing sequence.
Analogously, an axis $A$ can be transformed by at most $k$ swaps so that a vote $V$ is single-peaked if and only if there exists a permutation $\sigma$ with at most $k$ inversions such that $\sigma^{-1}$ applied to $p(A,V)$ consists of a decreasing sequence followed by an increasing sequence.
The statement of the lemma follows now from the well-known fact that the number of inversions in a permutation $\pi$ equals the number of inversions in $\pi^{-1}$ and, consequently, using the inverse permutations we can transform a series of swaps in $V$ to a series of swaps in $A$ -- and vice versa.
\end{proof}

\subsubsection*{\emph{$k$-Candidate Partition} (\dcp)}
The Candidate Partition distance is similar to Additional Axes: Whereas Additional Axes requires a partition of votes, now we partition the set of candidates.
For each candidate set in the partition, the profile restricted this candidate set has to be single-peaked consistent.
This notion is useful, for example, in the following situation.
Each candidate has an opinion on a controversial Yes/No-issue.
Depending on their own preference voters will always rank all Yes-candidates before or after all No-candidates.
It might be that when considering only the Yes- or only the No-candidates, the election is single-peaked.
Therefore, if we acknowledge the importance of this Yes/No-issue and partition the candidates accordingly, we may obtain two single-peaked elections.

\begin{definition}
\label{def:cp}
Let $C_1, \ldots, C_{k+1}$ be a partition of $C$.
A profile $\mathcal{P}$ is \emph{$k$-candidate partition single-peaked with respect to an axis $A$ and $C_1, \ldots, C_{k+1}$}
if the profiles $\mathcal{P}[C_1], \ldots, \mathcal{P}[C_{k+1}]$ are single-peaked with respect to $A$.
Furthermore, $\mathcal{P}$ is \emph{$k$-candidate partition single-peaked consistent}
if there exist an axis $A$ and a partition $C_1, \ldots, C_{k+1}$ of $C$ such that $\mathcal{P}$ is \emph{$k$-candidate partition single-peaked with respect to $A$ and $C_1, \ldots, C_{k+1}$}.
Let $\dcp(\mathcal{P})$ denote the smallest $k$ such that $\mathcal{P}$ is $k$-candidate partition single-peaked consistent.
\end{definition}

Note that $\dcp(\mathcal{P})\leq \frac{m}{2}$ always holds.

{\addtocounter{example}{-1}
\stepcounter{continueexcnt}%
\begin{example}[continued]
  We partition the candidates into $C_1=\{a,e\}$ and $C_2=\{b,c,d\}$.
  Notice that $\mathcal{P}[C_1]$ is trivially single-peaked consistent because it contains only two candidates.
  Furthermore, $\mathcal{P}[C_2]$ contains only votes of the form $b c d$ and $d c b$.
  Thus, $\dcp(\mathcal{P})=1$.
\end{example}}

We remark that $k$-candidate partition is related to the notion of \emph{$k$-peaked elections}, introduced by \citeA{atal/YangG14}. A profile $\mathcal{P}$ is $k$-peaked with respect to an axis $A$ if for every vote $\vote{}\in\mathcal{P}$ there exists a partition $C_1, \ldots, C_{k}$ of $C$ that yields single-peakedness of $\succ\![C_i]$ with respect to $A$ for every $i\in\{1,\dots,k\}$.
In this sense, the $k$-peaked distance can be considered a local variant of the $k$-candidate partition distance, as a different partition can be chosen for every vote.

\subsubsection*{\emph{$k$-Clones} (\dc)}
Elkind, Faliszewski, and Slinko~\citeyear{elk-fal-sli:c:clone-structures} studied \emph{clone sets} in elections. A clone set is a set of candidates that are ranked consecutively in every vote, but not necessarily in the same order. %
Clone sets can be used to obtain single-peaked profiles via \emph{decloning}, i.e., clone sets are replaced by a single candidate contained in this clone set. The distance to single-peakedness is here the minimal number of clones that need to be removed from the election via decloning in order to make it single-peaked.

\begin{definition}
\label{def:c}
We say that the profile $\mathcal{P}$ is \emph{$k$-clones single-peaked with respect to an axis $A$},
if we can obtain a set $C'\subseteq C$ by removing at most $k$ clones from $C$ via decloning such that the preference profile $\mathcal{P}[C']$ is single-peaked with respect to $A$. 
Furthermore, we say that the profile $\mathcal{P}$ is \emph{$k$-clones single-peaked consistent}
if there exists an axis $A$ such that $\mathcal{P}$ is \emph{$k$-clones single-peaked with respect to $A$}.
Let $\dc(\mathcal{P})$ denote the smallest $k$ such that $\mathcal{P}$ is $k$-clones single-peaked consistent.
\end{definition} 

Note that $\dc(\mathcal{P})\leq m-1$ always holds.

{\addtocounter{example}{-1}
\stepcounter{continueexcnt}%
\begin{example}[continued]
  In our example we can obtain single-peakedness by decloning $\{ a,b\}$ and $\{ d, e \}$.
  Since $\dcd(\mathcal{P})= 2$ and deleting candidates is more general than decloning, $\dc(\mathcal{P})$ can not be less than $2$. Thus, $\dc(\mathcal{P})=2$.
\end{example}}

\subsubsection*{\emph{$k$-Width} (\dw)}
\emph{Clustered single-peakedness}, as introduced by~\citeA{cor-gal-spa:c:single-peaked-width}, is a notion strongly related to the Clones measure (clone sets are called clusters in their paper). 
Given a partition of the candidates into clone sets such that the preferences are single-peaked after decloning, the \emph{width} of a partition is the size of the largest clone set minus one. Since there are several partitions of preferences into clone sets, the distance single-peaked \emph{Width} is defined as the minimum width among all possible partitions of candidates into clone sets. 

\begin{definition}
\label{def:w}
We say that the profile $\mathcal{P}$ is \emph{$k$-width single-peaked with respect to an axis $A$},
if we can obtain a partition of $C$ into clone sets $C_1, \ldots, C_{\ell}$ such that the size of the largest clone set is $k+1$ and the profile resulting from decloning is single-peaked with respect to~$A$.
Furthermore, we say that the profile $\mathcal{P}$ is \emph{$k$-width single-peaked consistent}
if there exists an axis $A$ such that $\mathcal{P}$ is \emph{$k$-width single-peaked with respect to $A$}.
Let $\dw(\mathcal{P})$ denote the smallest $k$ such that $\mathcal{P}$ is $k$-width single-peaked consistent.
\end{definition} 

Note that $\dw(\mathcal{P})\leq m-1$ always holds.

{\addtocounter{example}{-1}
\stepcounter{continueexcnt}%
\begin{example}[continued]
Again, partition $C$ into the clone sets $C_1=\{ a,b\}$, $C_2=\{ c\}$, and $C_3=\{ d, e \}$. The resulting decloned profile is single-peaked, the size of the largest clone set is two, and
thus $\dw(\mathcal{P})=1$.
\end{example}}

Another notion appearing in the literature is the Swoon distance introduced by \citeA{fal-hem-hem:j:nearly-single-peaked}.
A profile $\mathcal{P}$ is $(k,k')$-Swoon with respect to $A$ if by removing the top $k$ and the last $k'$ candidates from each vote yields a partial profile that is single-peaked with respect to $A$.
Due to the two parameters $k$ and $k'$, this notion does not immediately yield a clear definition of distance and hence we have excluded it from our study.
Note, however, that Local Candidate Deletion is a natural generalization of this concept.

\subsubsection*{\emph{Decision Problems}}
We now introduce the decision problems we will study. We define the following problems for $\mbox{X}\in \{$Voter Deletion, Candidate Deletion, Local Candidate Deletion, Additional Axes, Global Swaps, Local Swaps, Candidate Partition, Clones, Width$\}$.

\EP{X Single-Peaked Consistency}
{An election $E=(C,\mathcal{P})$ and a positive integer $k$.}
{Is $\mathcal{P}$ $k$-X single-peaked consistent?}

\EP{X Single-Peaked Evaluation}
{An election $E=(C,\mathcal{P})$, a positive integer $k$ and an axis\footnotemark $A$.}
{Is $\mathcal{P}$ $k$-X single-peaked with respect to $A$?}
\footnotetext{For Additional Axes we assume that $k+1$ axes $A_1, \ldots, A_{k+1}$ are given in the input (cf.\ Definition~\ref{def:aa}). For Candidate Partition we assume that an axis $A$ together with a partition $C_1, \ldots, C_k$ is given in the input (cf.\ Definition~\ref{def:cp}).}
The complexity of these problems has not been studied with the exception of \textsc{Clones Single-Peaked Consistency} and \textsc{Width Single-Peaked Consistency}, both of which are solvable in polynomial time \cite{elk-fal-sli:c:clone-structures,DBLP:conf/ijcai/CornazGS13}.
Furthermore, independent from our work, \citeA{bre-che-woe-profiles-nearby} have shown that \textsc{Voter Deletion Single-Peaked Consistency} is \np-complete.
In Section~\ref{sec:complexity-cons} we will study the complexity of these two decision problems in detail.

\section{Basic Results about Single-Peaked Profiles}
 \label{sec:results-basic}

In this section we collect simple facts about the single-peaked restriction, which we will use in several proofs.

\begin{lemma}
\label{lem:1..n,n..1}
Let $\mathcal{P}$ be a preference profile containing the vote $\succ: c_1 \ldots c_m$ and its reverse 
$\overline{\succ}$.
Then $\mathcal{P}$ is either single-peaked with respect to the axis $c_1>\cdots >c_m$ (and its reverse) or it is not single-peaked at all.
\end{lemma}

\begin{proof}
Since the vote $\vote{}$ ranks $c_m$ last while the vote 
$\overline{\vote{}}$ ranks $c_1$ last, these candidates have to be at the left-most and right-most position on any compatible axis.
Note that $c_1$ is the top-ranked candidate of $\vote{}$.
Hence this already determines the position of all other candidates.
Consequently only two axes are possible: $c_1>\cdots >c_m$ and $c_m>\cdots >c_1$.
\end{proof}

Lemma~\ref{lem:not-spc-crit} provides an alternative characterization of single-peakedness. 

\begin{lemma}\label{lem:not-spc-crit}
  Given an election $(C,\mathcal{P})$, the profile $\mathcal{P}$ is not single-peaked consistent with respect to an axis $A$ if and only if there is some voter ${\vote{}}\in\mathcal{P}$ and three candidates $c_i, c_j, c_k\in C$ such that $c_i>c_j>c_k$ on axis $A$, and $c_i\succ c_j$ holds as well as $c_k\succ c_j$.
\end{lemma}

\begin{proof}
  Assume that $\mathcal{P}$ is not single-peaked consistent.
  Then, for each axis $A$, there has to exist some voter $v$ that is not single-peaked with respect to $A$.
  Let $c$ be the top-ranked candidate of voter $v$.
  Then there exist candidates $c_1,c_2\in C$ with either $c>c_1>c_2$ or $c_2>c_1>c$ such that $c_2\succ c_1$.
  Depending on whether $c>c_1>c_2$ or $c_2>c_1>c$ we can instantiate $(c_i,c_j,c_k)$ with either $(c,c_1,c_2)$ or with $(c_2,c_1,c)$.
  It is now easy to see that $c_i>c_j>c_k$, $c_i\succ c_j$ and $c_k\succ c_j$.

  Let $A$ be an axis on $C$.
  For the converse direction assume that there is some voter $v$ and three candidates $c_i, c_j, c_k\in C$ such that $c_i>c_j>c_k$ on axis $A$, $c_i\succ c_j$ and $c_k\succ c_j$.
  Notice that $c_j$ cannot be the top-ranked candidate of voter $v$ as this would contradict $c_i\succ c_j$ and $c_k\succ c_j$.
  Thus, the top-ranked candidate $c$ lies either left or right of $c_j$ on axis $A$.
  We consider only the first case -- the other case can be dealt with analogously.
  It holds that $c>c_j>c_k$. 
  Definition~\ref{def:single-peaked} now requires $c\succ c_j \succ c_k$ for $v$ to be single-peaked with respect to $A$.
  This condition is, however, violated by our assumption $c_k\succ c_j$.
  Therefore $\mathcal{P}$ is not single-peaked consistent.
\end{proof}

The following observation says that any subelection of a single-peaked election is also single-peaked. 

\begin{lemma}\label{lem:spc-subsets}
 Let $(C,\mathcal{P})$ be a given election and $C'\subseteq C$.
 If $\mathcal{P}$ is single-peaked consistent then also $\mathcal{P}[C']$ is single-peaked consistent.
\end{lemma}

\begin{proof}
  Assume towards a contradiction that there is some $C'\subseteq C$ such that $\mathcal{P}[C']$ is not single-peaked consistent.
  Let $A$ be an arbitrary axis ordering $C$.
  By Lemma~\ref{lem:not-spc-crit} there is some voter ${\vote{}}\in\mathcal{P}$ and three candidates $c_i, c_j, c_k\in C'$ such that $c_i>c_j>c_k$ on the axis $A[C']$, $c_i\succ c_j$ and $c_k\succ c_j$.
  Then, however, it also holds that $c_i>c_j>c_k$ on the axis $A$ since $A$ is an extension of $A[C']$.
  Therefore the right-hand side of Lemma~\ref{lem:not-spc-crit} holds for every axis $A$ on $C$.
  Hence, by Lemma~\ref{lem:not-spc-crit}, $\mathcal{P}$ is not single-peaked consistent.
\end{proof}

The following lemma is an immediate consequence of the single-peaked classification theorem of \citeA{bal-hae:j:characterization-single-peaked}. For completeness we prove this much simpler statement directly.

\begin{lemma}
\label{lem:three-last-ranked}%
An election $(C,\mathcal{P})$ is not single-peaked if there exist three candidates $c_1,c_2,c_3\!\in\! C$ and three votes $V_1,V_2,V_3\in\mathcal{P}$ such that, for $i\in\{1,2,3\}$, $c_i$ is ranked last in $V_i[\{c_1,c_2,c_3\}]$.
\end{lemma}
\begin{proof}
It is straightforward to verify that for any axis $A$ on $\{c_1,c_2,c_3\}$ with candidate $c_i$ in the middle, vote $V_i$ is not single-peaked with respect to $A$. Hence, by Lemma~\ref{lem:spc-subsets}, $(C,\mathcal{P})$ is not single-peaked.
\end{proof}

\section{Relations between Notions of Nearly Single-Peakedness}
\label{sec:relations}

Theorem~\ref{thm:relations} states inequalities that hold for the distance measures under consideration.
We hereby show how these measures relate to each other.
For an overview consult Figure~\ref{fig:hasse}: In this Hasse diagram one distance measure $X$ is above and connected to a distance measure $Y$ if the measure $Y$ can be bounded by a function of measure $X$.
Intuitively, distances at the top are more fine-grained notions.
More formally, Figure~\ref{fig:hasse} displays a partial order defined as follows. For two distances measures $X$ and $Y$, $Y\leq X$ if and only if there exists a function $f$ such that $Y(\mathcal{P})\leq f(X(\mathcal{P}))$ for any profile $\mathcal{P}$.

\begin{theorem}
\label{thm:inequalities}
Let $(C,\mathcal{P})$ be an election. Then the following inequalities hold:
\begin{align*}
&\textit{(1)\ \ } \dls(\mathcal{P})\leq \dgs(\mathcal{P}). & & \textit{(5)\ \ } \dvd(\mathcal{P}) \leq \dgs(\mathcal{P}). & & \textit{(9)\ \ } \dcp(\mathcal{P}) \leq \dw(\mathcal{P}).\\
&\textit{(2)\ \ } \dlcd(\mathcal{P})\leq \dcd(\mathcal{P}). & & \textit{(6)\ \ } \daa(\mathcal{P}) \leq \dvd(\mathcal{P}). & & \textit{(10)\ \ } \dcd(\mathcal{P}) \leq \dc(\mathcal{P}).\\
&\textit{(3)\ \ } \dcd(\mathcal{P}) \leq \dgs(\mathcal{P}). & & \textit{(7)\ \ } \dcp(\mathcal{P}) \leq \dcd(\mathcal{P}). & & \textit{(11)\ \ } \dw(\mathcal{P}) \leq \dc(\mathcal{P}).\\
&\textit{(4)\ \ } \dlcd(\mathcal{P})\leq \dls(\mathcal{P}). & & \textit{(8)\ \ } \dcp(\mathcal{P}) \leq \dls(\mathcal{P}). & & \textit{(12)\ \ } \daa(\mathcal{P}) \leq (6\cdot\dc(\mathcal{P}))!.
\end{align*}
This list is complete in the following sense:
Inequalities that are not listed here and that do not follow from transitivity do not hold in general.
\label{thm:relations}
\end{theorem}
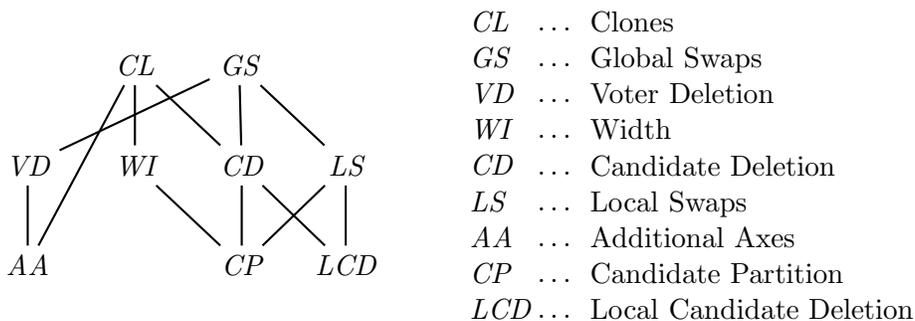
\begin{figure}[t]
\centering\begin{tikzpicture}[node distance=8mm and 6mm]
  \tikzstyle{nde} = [minimum size=0.0cm]
  \tikzstyle{box} = [draw=black,dashed]
  \node (c) {$\dc$};
  \node (gs) [right=of c] {$\dgs$};
  \node (w) [below=of c] {$\dw$};
  \node (m) [left=of w] {$\dvd$};
  \node (aa) [below=of m]{$\daa$};
  \node (cd) [right=of w] {$\dcd$};
  \node (cp) [below=of cd] {$\dcp$};
  \node (ls) [right=of cd] {$\dls$};
  \node (lcd) [below=of ls] {$\dlcd$};

  \foreach \from/ \to in {aa/m,m/gs,cp/cd,cd/gs,lcd/cd,lcd/ls,ls/gs,cp/ls,cp/w,w/c,aa/c,cd/c}
    \draw[draw=black,thick,-] (\from) -- (\to); %
    \node (legend1) at ($ (ls)+(2.1,0) $) [text width=0.9cm] {\dc\\\dgs\\\dvd\\\dw\\\dcd\\\dls\\\daa\\\dcp\\\dlcd};

    \node (legenddots) at ($ (legend1)+(0.7,-0.1) $) [text width=0.5cm] {$\ldots$\\$\ldots$\\$\ldots$\\$\ldots$\\$\ldots$\\$\ldots$\\$\ldots$\\$\ldots$\\$\ldots$};

    \node (legend2) at ($ (legend1)+(3.5,0) $) [text width=4.7cm] {Clones\\Global Swaps\\Voter Deletion\\Width\\Candidate Deletion\\Local Swaps\\Additional Axes\\Candidate Partition\\Local Candidate Deletion};
\end{tikzpicture}
\caption{The relation of distance measures (cf.\ Theorem~\ref{thm:inequalities}). This Hasse diagram shows a partial order defined as $Y\leq X$ if and only if $Y$ can be bounded by a function of $X$.}
\label{fig:hasse}
\end{figure}
\begin{proof}
Inequalities~1 and 2 are immediate consequences from the definitions since \dls permits more swaps than \dgs and \dlcd is more flexible than \dcd.
Inequalities~3 and 4 are due to the fact that swapping two candidates in a vote is at most as effective as removing one of these candidates.
Similarly, for Inequality~5 observe that removing the corresponding voter is at least as effective as swapping two candidates in the vote.
Concerning Inequality~6 observe that instead of deleting a voter we can always add an additional axis for this voter.
Inequality~7 follows from the fact that putting each deleted candidate in its own partition leads to single-peakedness if deleting these candidates does.

In order to show Inequality~8 let $\mathcal{P}$ be $k$-local swaps single-peaked consistent.
This means that there exists an axis $A$ such that after performing at most $k$ swaps per voter, $\mathcal{P}$ becomes single-peaked with respect to $A$.
Without loss of generality assume that the axis $A$ is $c_1 > c_2 > \cdots > c_m$.
We now partition the candidates in $k+1$ sets $S_0,\ldots,S_k$.
This is done by putting the $i$-th largest element of $A$ into the ($i$ modulo $k+1$)-th set.
Since we assume that $A$ is $c_1 > c_2 > \cdots > c_m$, we can equivalently say that $c_i$ is put into the ($i$ modulo $k+1$)-th set,
i.e., the $c_1$ in $S_1$, the $c_2$ in $S_2$, the $c_k$ in $S_k$ and $c_{k+1}$ in $S_0$.
Let $S\in\{S_0,\ldots, S_k\}$.
Towards a contradiction assume that $\mathcal{P}[S]$ is not single-peaked with respect to $A[S]$.
By Lemma~\ref{lem:not-spc-crit} there exists some voter ${\vote{}}\in\mathcal{P}$ and three candidates $c_x, c_y, c_z\in C$ with 
$x<y<z$, $c_x\succ c_y$ and $c_z\succ c_y$.
On axis $A$ the distance between $c_x$ and $c_y$ respectively $c_y$ and $c_z$ is at least $k+1$, i.e., at least $k$ elements lie in between them.
We know that at most $k$ swaps in $\vote{}$ can make this vote single-peaked with respect to $A$.
Let $\vote{}'$ denote this swapped vote.
Necessarily, these swaps have to either cause that $c_y\succ' c_{y-1}\succ'\cdots\succ' c_{x+1}\succ' c_x$ holds or that $c_y\succ'c_{y+1}\succ'\cdots\succ' c_{z-1}\succ' c_z$ holds in $\vote{}'$ (depending whether the top-ranked candidate of $\vote{}'$ is right or left of $c_y$).
Let us focus on the case that the swaps ensure that $c_y\succ' c_{y-1}\succ'\cdots\succ' c_{x+1}\succ' c_x$ -- the other case is analogous.
For $\vote{}$, contrary to $\vote{}'$, it holds that $c_x \succ c_y$.
Hence these swaps have to cause that $c_y \succ' c_x$ holds.
In addition, at least $k$ elements, namely $c_{x+1},\ldots, c_{y-1}$, have to be in between them. This requires at least $k+1$ swaps which contradicts the fact that at most $k$ swaps suffice.
Therefore, for all partition sets $S$, $\mathcal{P}[S]$ is single-peaked consistent and $\dcp(\mathcal{P}) \leq \dls(\mathcal{P})$.

To prove Inequality~9, consider a partition into clone sets. If we partition $C$ in sets so that every set contains at most one element from each clone set, then we obtain a partition consisting of $\dw(\mathcal{P})+1$ sets, since the original partition consisted of sets with at most $\dw(\mathcal{P})+1$ elements. The given profile is $\dw(\mathcal{P})$-candidate partition single-peaked with respect to this partition and hence $\dcp(\mathcal{P})\leq \dw(\mathcal{P})$.

Inequality~10 is an immediate consequence of the definitions since removing clones is a restricted form of deleting candidates. To see Inequality~11 note that $\dc$ is the total number of candidates removed via decloning whereas $\dw$ only measures the number of candidates removed in the largest clone set.

For Inequality~12,
let $C'$ be the reduced candidate set obtained by decloning and let $A'$ be an axis such that $\mathcal{P}[C']$ is single-peaked with respect to $A'$.
We will show that by building upon $A'$ we can construct an axis compatible with all votes that order certain candidates (namely $D_1\cup D_2$, as defined below) in the same way.
Let $D_1$ be the set of all candidates appearing in clone sets of size at least $2$ and let $D_2$ be the set of all right and left neighbors on $A'$ of candidates in $D_1\cap C'$.
To bound the size of $D_1$ observe that in every clone set of size $c$ exactly $c-1$ candidates are decloned and for $c\geq 2$ it holds that $c\leq 2(c-1)$. Thus, we can bound $|D_1|\leq 2\cdot\dc(\mathcal{P})$.
Since $D_2$ contains at most twice as many candidates as $D_1$, it holds that $|D_2|\leq 4\cdot\dc(\mathcal{P})$.
Hence, $|D_1\cup D_2|\leq 6\cdot\dc(\mathcal{P})$ and the total number of permutations on $D_1\cup D_2$ is bounded by $(6\cdot\dc(\mathcal{P}))!$.

For each possible permutation of $D_1\cup D_2$ we consider a separate axis, i.e., two votes share the same axis only if they agree on the order of all candidates in $D_1\cup D_2$.
Let $T$ be a fixed order on $D_1\cup D_2$; we write $a\succ_T b$ to compare two candidates with respect to $T$.
We intend to build an axis that is compatible with all votes agreeing with $T$ on the order of candidates in $D_1\cup D_2$.
This is done as follows:
Let $C''$ be a clone set and $d$ its representative on $A'$, i.e., $d$ is the unique candidate in $C''$ that has not been decloned.
Further let $c$ and $e$ be the elements left and right of $d$ on $A'$, respectively.
We replace the contiguous sequence $c>d>e$ of $A'$ with $c> T[C'] > e$ if $c\succ_T d$ and with $c> \overline{T[C']} > e$ if $d\succ_T c$, i.e., the elements of $C'$ are ordered descending (ascending) according to $T$ if $c$ is larger (smaller) than $d$ in all votes compatible with $T$.
We repeat this procedure for all clone sets and obtain an axis $A$ for $C$ that witnesses single-peakedness for all votes compatible with $T$.

To see that this is the case, we employ Lemma~\ref{lem:not-spc-crit}: Assume towards a contradiction that there is some voter ${\vote{}}$ in $\mathcal{P}$ and three candidates $c_i, c_j, c_k\in C$ such that $c_i>c_j>c_k$ on axis $A$, and $c_i\succ c_j$ holds as well as $c_k\succ c_j$.
If the representatives of $c_i$, $c_j$, and $c_k$ are all different, then also the representatives violate the single-peaked condition, which contradicts our assumption that $A'$ is a single-peaked axis for all representatives.
If $c_i$, $c_j$, and $c_k$ all have the same representative, then we obtain a contradiction to our assumption that $c_i\succ c_j$ and $c_k\succ c_j$ from the fact that either $c_i\succ c_j\succ c_k$ (or the reverse) holds in all votes compatible with $T$.
If $c_i$, $c_j$, and $c_k$ have two representatives in total, we can assume without loss of generality that $c_j$ and $c_k$ share the same representative (i.e., they are in the same clone set), since if $c_i$ and $c_k$ are in the same clone set then also $c_j$ is in this clone set.
(The case where $c_i$ and $c_j$ share the same representative is analogous.)
Observe that $\{c_j, c_k\}\subseteq D_1$ and hence $c_i\in D_2$. Thus the order of these three candidates is determined by $T$ and therefore the same in all votes under consideration. Furthermore, $c_j$ and $c_k$ are consecutive in $T$ since they are in the same clone set and hence, due to our assumption that $c_i \succ c_j$ and $c_k\succ c_j$, the only possibility to order the three candidates is $c_i\succ c_k\succ c_j$. This, however, implies that either $c_i> c_k> c_j$ or the reverse holds on $A$ by our construction, a contradiction. 
Therefore, we conclude that all votes compatible with $T$ are single-peaked with respect to our constructed axis $A$, and consequently, we can bound $\daa(\mathcal{P}) \leq (6\cdot\dc(\mathcal{P}))!$.

It remains to show that the inequalities listed in the theorem statement are indeed all inequalities that hold for these measures.
To this end we provide counterexamples for each remaining case of the following form: To show that measure X cannot be bounded measure Y, we present a sequence of elections with arbitrary large X and constant Y.
In these sequences of elections either the number of votes $n$ or the number of candidates $m$ (or both) is growing.
Table~\ref{tab:inequalities} offers an overview by pointing to the corresponding counterexample.

\begin{table}[t]
  \centering
  \begin{tabular}{r||c|c|c|c|c|c|c|c|c|}
    \multicolumn{1}{r||}{} & \begin{sideways}\dvd$(\mathcal{P})$\end{sideways} & \begin{sideways}\dcd$(\mathcal{P})$\end{sideways} & \begin{sideways}\dlcd$(\mathcal{P})$\end{sideways} & \begin{sideways}\dgs$(\mathcal{P})$\end{sideways} & \begin{sideways}\dls$(\mathcal{P})$\end{sideways} & \begin{sideways}\daa$(\mathcal{P})$\end{sideways} & \begin{sideways}\dcp$(\mathcal{P})$\end{sideways} & \begin{sideways}\dc$(\mathcal{P})$\end{sideways} & \begin{sideways}\dw$(\mathcal{P})$\end{sideways}\\\hline\hline
    Voter Deletion\dotfill \dvd$(\mathcal{P})$& = & \ref{cex:1} & \ref{cex:3}& $\leq$ &\ref{cex:3} & \ref{cex:1}& \ref{cex:1} & \ref{cex:16} & \ref{cex:16} \\ \hline
    Candidate Deletion\dotfill \dcd$(\mathcal{P})$ & \ref{cex:2}& =& \ref{cex:3}& $\leq$ & \ref{cex:3} & \ref{cex:2}&\ref{cex:2} & $\leq$ & \ref{cex:12}  \\ \hline
    Local Candidate Deletion $\ldots$ \dlcd$(\mathcal{P})$ & \ref{cex:2}& $\leq$ &= &$\leq$ & $\leq$ & \ref{cex:2}&\ref{cex:2} & $\leq$ & \ref{cex:12} \\ \hline
    Global Swaps\dotfill \dgs$(\mathcal{P})$ & \ref{cex:7}& \ref{cex:7}& \ref{cex:7} &= &\ref{cex:3} &\ref{cex:7} &\ref{cex:7} & \ref{cex:16} & \ref{cex:16}  \\ \hline
    Local Swaps\dotfill \dls$(\mathcal{P})$ & \ref{cex:7} &\ref{cex:7} & \ref{cex:7}& $\leq$ &= &\ref{cex:7} &  \ref{cex:7} &  $\leq$ & \ref{cex:12} \\ \hline
    Additional Axes\dotfill \daa$(\mathcal{P})$ & $\leq$&\ref{cex:8} &\ref{cex:8} &$\leq$ & \ref{cex:11} &= & \ref{cex:8} & $\leq$ & \ref{cex:12}  \\ \hline
    Candidate Partition\dotfill \dcp$(\mathcal{P})$ & \ref{cex:10}&$\leq$ & \ref{cex:4}& $\leq$ &$\leq$ &\ref{cex:10} &= & $\leq$ & $\leq$\\ \hline
    Clones\dotfill \dc$(\mathcal{P})$ & \ref{cex:9}&  \ref{cex:9}  & \ref{cex:9}&  \ref{cex:9}  &  \ref{cex:9}  &\ref{cex:9} &  \ref{cex:9} & = & \ref{cex:13} \\ \hline
    Width\dotfill \dw$(\mathcal{P})$ & \ref{cex:9}&   \ref{cex:9} & \ref{cex:9}&  \ref{cex:9} &  \ref{cex:9} & \ref{cex:9} &  \ref{cex:9} & $\leq$ & = \\ \hline

  \end{tabular}
  \caption{Inequalities regarding the distance measures. This table should be read as follows. Measures in the left-most column are bounded ($\leq$) by a function of the measures in the top row. Numbers point to the corresponding counterexample if no such bound exists.}
  \label{tab:inequalities}
\end{table}

\medskip\noindent\counterex{cex:1}{\dvd cannot be bounded by \dcd, \daa and \dcp}
Consider the preference profile on the candidate set $C=\{c_1,\ldots,c_m\}$ with the following $2m$ votes:
\begin{itemize}
\item There are $m$ votes of the form:
$\begin{array}{c@{\ \ }c@{\ \ }c@{\ \ }c}
c_1 & c_2 & \ldots & c_m.
\end{array}$
\item There are $m$ votes of the form:
$\begin{array}{c@{\ \ }c@{\ \ }c@{\ \ }c@{\ \ }c@{\ \ }c}
c_m & c_2 & c_3 & \ldots & c_{m-1} & c_1.
\end{array}$
\end{itemize}

The corresponding preference profile $\mathcal{P}$ is not single-peaked consistent.
This is because $c_2$ has to be next to both $c_1$ and $c_m$ on any suitable axis but both have to be either the left-most or right-most element.
Consequently, $\dvd(\mathcal{P})=m$.
Removing candidates instead of voters is far more useful in this case.
When we remove either $c_1$ or $c_m$, $\mathcal{P}$ becomes single-peaked and hence $\dcd(\mathcal{P})=1$.
Since we have only two distinct votes, we require two axes to make $\mathcal{P}$ single-peaked and hence $\daa(\mathcal{P})=1$.
Furthermore, notice that we can obtain single-peaked consistency by partitioning the candidates into two sets $C_1=\{c_1,c_m\}$ and $C_2=\{c_2,\ldots,c_{m-1}\}$, hence $\dcp(\mathcal{P})=1$.

\medskip\noindent\counterex{cex:7}{Neither \dgs nor \dls can be bounded by \dvd, \daa, \dcd, \dlcd and \dcp}
This counterexample is similar to the previous one but $\mathcal{P}$ consists of only two votes.
Let $C=\{c_1,\ldots,c_{3m+1}\}$ be the set of candidates.
\begin{itemize}
\item There is one vote of the form:
$\begin{array}{c@{\ \ }c@{\ \ }c@{\ \ }c}
c_1 & c_2 & \ldots & c_{3m+1}.
\end{array}$
\item There is one vote of the form:
$\begin{array}{c@{\ \ }c@{\ \ }c@{\ \ }c@{\ \ }c@{\ \ }c}
c_{3m+1} & c_2 & c_3 & \ldots & c_{3m} & c_1.
\end{array}$
\end{itemize}

If we consider $\mathcal{P}\left[\{c_1,c_{m+1},c_{2m+1},c_{3m+1}\}\right]$, we observe that this restricted profile is not single-peaked.
Consequently, by Lemma~\ref{lem:spc-subsets}, $\mathcal{P}$ is not single-peaked as well.
If we want to make $\mathcal{P}$ single-peaked via swaps, at least two of $\{c_1,c_{m+1},c_{2m+1},c_{3m+1}\}$ have to swap position.
This requires at least $m$ swaps and consequently $\dgs(\mathcal{P})\geq \dls(\mathcal{P})\geq m$.
Since there are only two votes, $\daa(\mathcal{P})=\dvd(\mathcal{P})=1$.
As in the previous counterexample removing either $c_1$ or $c_{3m+1}$ yields a single-peaked profile and hence $\dlcd(\mathcal{P})=\dcd(\mathcal{P})=\dcp(\mathcal{P})=1$.

\medskip\noindent\counterex{cex:2}{Neither \dcd nor \dlcd can be bounded by \dvd, \daa and \dcp}
  This time we consider three votes on the candidates $C=\{ c_1, \ldots ,c_{2m} \}$.

\begin{itemize}
\item There is one vote $\vote{1}$ of the form:
$\begin{array}{c@{\ \ }c@{\ \ }c@{\ \ }c}
c_1 & c_2 & \ldots & c_{2m}.
\end{array}$
\item There is one vote $\vote{2}$ of the form:
$\begin{array}{c@{\ \ }c@{\ \ }c@{\ \ }c}
c_{2m} & c_{2m-1} & \ldots & c_1.
\end{array}$
\item There is one vote $\vote{3}$ of the form:
$\begin{array}{c@{\ \ }c@{\ \ }c@{\ \ }c@{\ \ }c@{\ \ }c}
c_m & \ldots & c_1 & c_{m+1} & \ldots & c_{2m}.
\end{array}$
\end{itemize}

By Lemma~\ref{lem:1..n,n..1} we only have to consider the axis $c_1>c_2>\cdots>c_{2m}$ for $\mathcal{P}=(\vote{1},\vote{2},\vote{3}\nobreak)$.
The third vote $\vote{3}$ is however not single-peaked with respect to this axis.
  Hence $\dvd(\mathcal{P})=\daa(\mathcal{P})=1$.
  Also, we have that $\dcp(\mathcal{P})=1$ since $\mathcal{P}[\{c_1,\ldots,c_m\}]$ and $\mathcal{P}[\{c_{m+1},\ldots,c_{2m}\}]$ are single-peaked consistent.
  However, we have to remove a lot of candidates to obtain a single-peaked profile.
  Indeed, we have to remove candidates such that the indices of the remaining candidates in $\vote{3}$ are either increasing or decreasing.
  That are at least $m-1$ to remove and hence $\dcd(\mathcal{P})\geq \dlcd(\mathcal{P})\geq m-1$.

\medskip\noindent\counterex{cex:3}{Neither \dvd, \dgs nor \dcd can be bounded by \dlcd and \dls}
  We consider an election with $3n$ votes on the candidates $C=\{ c_1, \ldots ,c_{3n} \}$.

\begin{itemize}
\item There are $n$ votes $\vote{1}, \ldots , \vote{n}$ of the form:
$\begin{array}{c@{\ \ }c@{\ \ }c@{\ \ }c}
c_1 & c_2 & \ldots & c_{3n}.
\end{array}$
\item There  are $n$ votes $\vote{n+1}, \ldots , \vote{2n}$ of the form:
$\begin{array}{c@{\ \ }c@{\ \ }c@{\ \ }c}
c_{3n} & c_{3n-1} & \ldots & c_1.
\end{array}$
\item  The remaining votes are obtained from the first vote by swapping the last two candidates in the $i$-th block consisting of three candidates. Formally, for each $i$, $1\leq i \leq n$ there is a vote $\vote{2n+i}$ of the form:
$c_1 \  \ldots \ c_{3(i-1)-1} \ \underbrace{c_{3(i-1)} \ c_{3(i-1)+2} \ c_{3(i-1)+1}}_{i\text{-th block}} \ c_{3i} \  \ldots \ c_{3n}$.
\end{itemize}

  Let $\mathcal{P}=(\vote{1},\vote{2},\ldots,\vote{3n})$.
  By using Lemma~\ref{lem:three-last-ranked} it is easy to check that for each $1 \leq i\leq n$, $\mathcal{P}[\{c_{3(i-1)+2},c_{3(i-1)+1},c_{3i}\}]$ is not single-peaked consistent.
  By Lemma~\ref{lem:spc-subsets}, $\mathcal{P}$ is not single-peaked consistent.
  Also, this implies that we have to remove at least one candidate in each set $\{c_{3(i-1)+2},c_{3(i-1)+1},c_{3i}\}$ in order to make $\mathcal{P}$ single-peaked consistent.
  Therefore $\dcd(\mathcal{P})\geq n$.
  Since $\dgs(\mathcal{P})\geq \dcd(\mathcal{P})$ also $\dgs(\mathcal{P})\geq n$.
  We now want to prove a lower bound on $\dvd(\mathcal{P})$.
  If we delete $n-1$ votes then at least one vote of $\{\vote{1},\ldots,\vote{n}\}$, one of $\{\vote{n+1},\ldots,\vote{2n}\}$ and one of $\{\vote{2n+1},\ldots,\vote{3n}\}$ remains.
  Again by Lemma~\ref{lem:spc-subsets} and \ref{lem:three-last-ranked}, $\mathcal{P}$ is not single-peaked consistent.
  Hence $\dvd(\mathcal{P})> n-1$.
  Finally, notice that the votes $\vote{2n+1},\ldots,\vote{3n}$ can be turned into vote $\vote{1}$ by a single swap, which shows that $\dls(\mathcal{P})=1$.
  Since $\dlcd(\mathcal{P})\leq \dls(\mathcal{P})$ also $\dlcd(\mathcal{P})=1$.

\medskip\noindent\counterex{cex:8}{\daa cannot be bounded by \dcd, \dlcd and \dcp}
In this example we use $n$ votes on the candidates $C=\{ c_1, \ldots ,c_{n+1} \}$.

\begin{itemize}
\item For each $i$, $1\leq i \leq n$, there is one vote $\vote{i}$ of the form:\\
$\begin{array}{c@{\ \ }c@{\ \ }c@{\ \ }c@{\ \ }c@{\ \ }c@{\ \ }c@{\ \ }c@{\ \ }c}
c_{n+1} & c_i & c_{i-1}  & \ldots & c_1 & c_{i+1} & c_{i+2} &  \ldots & c_{n}.
\end{array}$
\end{itemize}

Let us consider the preference profile $\mathcal{P}=(\vote{1},\vote{2},\ldots,\vote{n})$.
All votes have the same top-ranked candidate but different candidates in the second place.
If this preference profile was single-peaked then these second-place candidates had to be either left or right of the peak.
This is not possible for three or more candidates.
Hence the profile $\mathcal{P}$ containing three or more votes is not single-peaked.
Thus, $\daa(\mathcal{P})\geq \frac n 3-1$.
Deleting $c_{n+1}$ however makes $\mathcal{P}$ single-peaked with respect to the axis $c_1>c_2>\cdots>c_n$ and hence $\dcd(\mathcal{P})=\dlcd(\mathcal{P})=\dcp(\mathcal{P})=1$.

\medskip\noindent\counterex{cex:11}{\daa cannot be bounded by \dls}
We consider $n$ votes on $4n$ candidates $C=\{ c_1, \ldots ,c_{4n} \}$.

\begin{itemize}
\item For each $i$, $1\leq i \leq n$, there is one vote $\vote{i}$ of the form:\\
$\begin{array}{c@{\ \ }c@{\ \ }c@{\ \ }c@{\ \ }c@{\ \ }c@{\ \ }c@{\ \ }c@{\ \ }c@{\ \ }c}
c_1 & \ldots & c_{4i-4} & c_{4i}  & c_{4i-2} &  c_{4i-1} &  c_{4i-3} &  c_{4i+1} &  \ldots & c_{4n}.
\end{array}$
\end{itemize}

Let $\mathcal{P}=(\vote{1},\ldots,\vote{n})$.
The preference profile $\mathcal{P}$ is not single-peaked consistent since the restricted profiles $\mathcal{P}[\{c_{i-3},c_{i-2},c_{i-1},c_{i}\}]$, $i\in\{1,\ldots,n\}$, are neither.
With five swaps in each vote we can make these votes identical and hence $\dls(\mathcal{P})\leq 5$.
Even any pair of votes in $\mathcal{P}$ is not single-peaked.
Hence $\daa(\mathcal{P})\geq\frac n 2$.

\medskip\noindent\counterex{cex:4}{\dcp cannot be bounded by \dlcd}
Consider an election with $3n$ votes on the candidates $C=\{ c_1, \ldots ,c_{3n} \}$.

\begin{itemize}
\item For each $i$, $1\leq i \leq 3n$, there is one vote $\vote{i}$ of the form:\\
$\begin{array}{c@{\ \ }c@{\ \ }c@{\ \ }c@{\ \ }c@{\ \ }c@{\ \ }c@{\ \ }c}
c_1 & \ldots & c_{i-1} & c_{i+1}  & \ldots &  c_{3n} & c_i.
\end{array}$
\end{itemize}

Let $\mathcal{P}=(\vote{1},\ldots,\vote{3n})$.
Since the lowest ranked candidates have to be either at the left-most or right-most position on the axis and there are more than two lowest ranked candidates, this profile is not single-peaked consistent.
However, when the last-ranked candidate is removed in each vote, the profile becomes single-peaked consistent and hence $\dlcd(\mathcal{P})=1$.
Concerning $\dcp(\mathcal{P})$, notice that any partition into $n$ sets contains a set with at least three candidates and thus, by Lemma~\ref{lem:three-last-ranked} does not yield a single-peakedness.
Hence $n$ candidate partitions are not enough to obtain single-peaked consistency and hence $\dcp(\mathcal{P})\geq n$.

\medskip\noindent\counterex{cex:10}{\dcp cannot be bounded by \dvd and \daa}
Consider the candidate set $C=\{c_1,\ldots$, $ c_{m^2}\}$ and the following three votes:

\begin{itemize}
\item There is one vote $\vote{1}$ of the form:
$\begin{array}{c@{\ \ }c@{\ \ }c@{\ \ }c}
c_1 & c_2 & \ldots & c_{m^2}.
\end{array}$
\item There is one vote $\vote{2}$ of the form:
$\begin{array}{c@{\ \ }c@{\ \ }c@{\ \ }c}
c_{m^2} & c_{m^2-1} & \ldots & c_1.
\end{array}$
\item There is one vote $\vote{3}$ of the form:\\
$\begin{array}{c@{\ }c@{\ }c@{\ }c@{\ }c@{\ \ }c@{\ }c@{\ }c@{\ }c@{\ \ }c@{\ \ }c@{\ }c@{\ }c@{\ }c@{\ }c@{\ }c}
 c_1 & c_{m+1} & c_{2m+1} & \ldots & c_{m(m-1)+1} & c_{2} & c_{m+2} &  \ldots & c_{m(m-1)+2} & \ldots  & c_m & c_{2m} & c_{3m} & \ldots & c_{m^2}. 
\end{array}$
\end{itemize}

Let $\mathcal{P}=(\vote{1},\vote{2},\vote{3})$.
This preference profile is not single-peaked but $\dvd(\mathcal{P})=1$ and $\daa(\mathcal{P})=1$.
The candidates, however, have to be partitioned into many sets in order to obtain single-peakedness.
First, observe that by Lemma~\ref{lem:1..n,n..1} we only have to consider the axis $c_1>c_2>\cdots>c_{m^2}$.
Let us now consider vote $\vote{3}$.
Since we have fixed an axis we can consider longest increasing and decreasing subsequences in this vote.
Note that both increasing and decreasing subsequences have a length of less than $2m$.
Hence a subset of the candidates cannot be single-peaked if it contains more than $4m$ candidates.
We therefore have to partition the candidates of $\mathcal{P}$ into sets of cardinality at most $4m$ and by that $\dcp(\mathcal{P})\geq \frac m 4 - 1$.

\medskip\noindent\counterex{cex:16}{\dvd and \dgs cannot be bounded by \dc and \dw}
This counterexample uses $3n$ votes and three candidates.

\begin{itemize}
\item There are $n$ votes are of the form:
$\begin{array}{c@{\ \ }c@{\ \ }c}
c_1 & c_2 & c_3.
\end{array}$
\item There are $n$ votes are of the form:
$\begin{array}{c@{\ \ }c@{\ \ }c}
c_1 & c_3 & c_2.
\end{array}$
\item There are $n$ votes are of the form:
$\begin{array}{c@{\ \ }c@{\ \ }c}
c_2 & c_3 & c_1.
\end{array}$
\end{itemize}

Since all three candidates appear at the last position in votes, $n$ votes have to be deleted to make this profile single-peaked.
Analogously, at least $n$ swaps have to be performed.
Since $\{c_2,c_3\}$ can be decloned, \dc and \dw can be bounded by $1$.

\medskip\noindent\counterex{cex:9}{Neither \dw nor \dc can be bounded by \dgs, \dls, \dcp, \dcd, \dvd, \daa, and \dlcd}
Consider an election with four votes and $m$ candidates.

\begin{itemize}
\item There is one vote $\vote{1}$ of the form:
$\begin{array}{c@{\ \ }c@{\ \ }c@{\ \ }c}
c_1 & c_2 & \ldots & c_{m}.
\end{array}$
\item There is one vote $\vote{2}$ of the form:
$\begin{array}{c@{\ \ }c@{\ \ }c@{\ \ }c}
c_m & c_{m-1} & \ldots & c_{1}.
\end{array}$
\item There is one vote $\vote{3}$ of the form:
$c_1\ c_2\ \ldots\ c_{m-2}\ c_{m}\ c_{m-1} $.
\item There is one vote $\vote{4}$ of the form:
$c_{m-1}\ c_{m-2}\ \ldots\ c_1\ c_{m}$.
\end{itemize}

Let $\mathcal{P}=(\vote{1},\ldots,\vote{4})$.
Since $\mathcal{P}$ has three distinct last-ranked candidates, it is not single-peaked. Swapping candidates $c_{m-1}$ and $c_m$ in $\vote{3}$ provides us with a single-peaked profile according to axis $c_1 > c_2 > \cdots > c_m$,
thus $\dgs(\mathcal{P})=1$.
Let us consider \dc and \dw.
There are three last-ranked candidates: $c_1, c_{m-1}$ and $c_m$.
At least two of them have to be contained in a clone set.
As can easily be verified, such a clone set would have to be of size at least $m$.
Hence, $\dc(\mathcal{P})\geq\dw(\mathcal{P})\geq m-1$.
Since \dgs is an upper bound for \dls, \dcp, \dcd, \dvd, \daa, and \dlcd, none of them can bound \dw or \dc. 

\medskip\noindent\counterex{cex:12}{Neither \dcd,  \dlcd, \dgs nor \dls  can be bounded by \dw}
We consider an election with three votes and $2m+1$ candidates.

\begin{itemize}
\item There is one vote $\vote{1}$ of the form:
$\begin{array}{c@{\ \ }c@{\ \ }c@{\ \ }c}
c_1 & c_2 & \ldots & c_{2m+1}.
\end{array}$
\item There is one vote $\vote{2}$ of the form:
$\begin{array}{c@{\ \ }c@{\ \ }c@{\ \ }c}
c_{2m+1} & c_{2m} & \ldots & c_{1}.
\end{array}$
\item There is one vote $\vote{3}$ of the form:
$c_2\ c_1\ c_4\ c_3\ \ldots\ c_{2m-2}\ c_{2m-3}\ c_{2m}\ c_{2m-1} \ c_{2m+1} $.
\end{itemize}

Every candidate with an odd index is part of a valley, i.e., with respect to the axis fixed by $\vote{1}$ and $\vote{2}$ all candidates with odd indices are ranked below their two neighbors with respect to $\vote{3}$.
Hence $\dlcd$ is at least $m$ and by the Inequalities~2, 3 and 4 so are \dcd, \dgs and \dls.
On the contrary, if we put, for all $i\in \{1,\ldots,m\}$, $c_{2i}$ and $c_{2i-1}$ in a clone set, we obtain a single-peaked profile.
Since all these clone sets are of size $2$, $\dw$ is bounded~by~$1$.

\medskip\noindent\counterex{cex:13}{\dc cannot be bounded by \dw}
Let $C=\{c_1,\ldots,c_{3m}\}$. We define $3^m$ votes that are contained in $\mathcal{P}$.
For every sequence $s$ of length $m$ on the alphabet $\{X,Y,Z\}$ we define a vote.
The $i$-th entry of $s$ determines the order of $c_{3i-2},c_{3i-1},c_{3i}$.
If $s[i]=X$ then we choose the order $c_{3i-2} \succ c_{3i-1} \succ c_{3i}$.
If $s[i]=Y$ then we choose the order $c_{3i-2} \succ c_{3i} \succ c_{3i-1}$.
If $s[i]=Z$ then we choose the order $c_{3i-1} \succ c_{3i} \succ c_{3i-2}$.
The vote corresponding to $s$ is now defined as \[\{c_1,c_2,c_3\}\succ\{c_4,c_5,c_6\}\succ\cdots\succ\{c_{3m-2}, c_{3m-1}, c_{3m}\}\]
and the order in these sets is given by the sequence $s$ and the rules as defined above.

We have to remove $m$ candidates to make this profile single-peaked, i.e., $\dcd(\mathcal{P})=m$.
By Inequality~10, $\dc(\mathcal{P})\geq m$.
In contrast, grouping $c_{3i-2},c_{3i-1},c_{3i}$ together in clone sets (for all $i\in\{1,\ldots,m\}$) and decloning accordingly yields a single-peaked profile and hence $\dw(\mathcal{P})$ is bounded by $2$.
\end{proof}

We conclude this section by illustrating how the inequalities stated in Theorem~\ref{thm:inequalities} can be used to obtain new results.
More specifically, we will show that there are preference profiles that are close to being single-peaked but do not have a weak Condorcet winner -- in contrast to single-peaked profiles for which a weak Condorcet winner is guaranteed.
A \emph{weak Condorcet winner} is a candidate that is preferred to each other candidate by at least half of the voters.

\begin{proposition}\label{prop:no-condorcet-sp}
  For every $m\geq 3$ and $n\geq 1$ there is an election $E=(C,\mathcal{P})$ with $2n+1$ votes and $m$ candidates such that $\dgs(\mathcal{P})=1$, $\dc(\mathcal{P})=\dw(\mathcal{P})=2$ and $\mathcal{P}$ does not have a weak Condorcet winner.
\end{proposition}
\begin{proof}
  Let the set of candidates be $C=\{a,b,c\}\cup \{d_1,\ldots,d_{m-3}\}$.
  The profile $\mathcal{P}$ contains the following votes:
  \begin{itemize}
    \item a single vote of the form:
$\begin{array}{c@{\ \ }c@{\ \ }c@{\ \ }c@{\ \ }c@{\ \ }c}
b & c & a & d_1 & \ldots & d_{m-3},
\end{array}$
    \item $n$ votes of the form:
$\begin{array}{c@{\ \ }c@{\ \ }c@{\ \ }c@{\ \ }c@{\ \ }c}
a & b & c & d_1 & \ldots & d_{m-3}\text{, and}
\end{array}$
    \item $n$ votes of the form:
$\begin{array}{c@{\ \ }c@{\ \ }c@{\ \ }c@{\ \ }c@{\ \ }c}
c & a & b & d_1 & \ldots & d_{m-3}.
\end{array}$
  \end{itemize}
  It is straightforward to verify that the profile $\mathcal{P}$ does not have a weak Condorcet winner since it contains a cycle on $a,b,c$.
  Notice that $\mathcal{P}$ becomes single-peaked with respect to axis $b>a>c>d_1>\cdots>d_{m-3}$ if we swap candidates $b$ and $c$ in the first vote.
  Hence, we know that $\dgs(\mathcal{P})=1$.
  Furthermore, we obtain a single-peaked profile via decloning the clone set $\{a,b,c\}$ and consequently $\dc(\mathcal{P})=\dw(\mathcal{P})=2$.
\end{proof}

Due to the inequalities stated in Theorem~\ref{thm:inequalities}, the result of Proposition~\ref{prop:no-condorcet-sp} holds also if $\dgs(\mathcal{P})$ is replaced by one of the measures \dvd, \dcd, \dlcd, \dls, \daa, and \dcp.
(In principle these measures could be smaller than \dgs but the profile is not single-peaked and the distance is only $1$.)
Therefore, even a distance of $1$ to single-peakedness (with respect to the \dvd, \dcd, \dlcd, \dls, \daa, \dcp, and \dgs) does not help to avoid the Condorcet paradox.

Let us now briefly consider profiles with $\dc(\mathcal{P})=1$ or $\dw(\mathcal{P})=1$.
In contrast to our previous result, such profiles preserve an important property of single-peaked profiles, namely that weak Condorcet winners are guaranteed.

\begin{proposition}\label{prop:always-condorcet}
An election $E=(C,\mathcal{P})$ with $\dc(\mathcal{P})=1$ or $\dw(\mathcal{P})=1$ has a weak Condorcet winner.
\end{proposition}
\begin{proof}
If this statement holds for $\dw(\mathcal{P})=1$, it also holds for $\dc(\mathcal{P})=1$ since all profiles with $\dc(\mathcal{P})=1$ satisfy $\dw(\mathcal{P})=1$.
Let $C_1,\ldots,C_k$ be a partition of candidates into clone sets so that decloning these sets yields a single-peaked profile.
Let $C'$ be the corresponding set of decloned candidates, i.e., $\mathcal{P}[C']$ is single-peaked.
Let $w$ be a weak Condorcet winner of $\mathcal{P}[C']$ and let $w\in C_i$.
Observe that $\card{C_i}\leq 2$ since $\dw{\mathcal{P}}=1$.
For all candidates $c\in C\setminus C_i$ and all candidates $d\in C_i$, $d$ is preferred to $c$ by at least half the voters of $\mathcal{P}$.
If $C_i= \{w\}$, $w$ is a weak Condorcet winner in $\mathcal{P}$.
If $\card{C_i}= 2$ and the number of voters is odd, one of these two candidates is preferred over the other by more than half of the voters in $\mathcal{P}$; hence this preferred candidate is a weak Condorcet winner. If $\card{C_i}= 2$ and the number of voters is even, either one of these two candidates is preferred over the other by more than half of the voters in $\mathcal{P}$ and thus is a weak Condorcet winner or these two candidates are preferred over each other by exactly half of the voters, hence they are both weak Condorcet winners.
\end{proof}

\section{Computational Results}
\label{sec:complexity-cons}

In this section we study the complexity of \textsc{X Single-Peaked Consistency} and \textsc{X Single-Peaked Evaluation} for $\mbox{X}\in \{$Voter Deletion, Candidate Deletion, Local Candidate Deletion, Additional Axes, Global Swaps, Local Swaps, Candidate Partition$\}$.
The general theme is that \textsc{X Single-Peaked Consistency} is \np-complete whereas \textsc{X Single-Peaked Evaluation} is solvable in polynomial time.
The exception is the Candidate Deletion distance, for which also the consistency problem requires only polynomial time.
We do not consider consistency problems corresponding to the Clones and Width distance as it was already established that these are solvable in polynomial time \cite{elk-fal-sli:c:clone-structures,DBLP:conf/ijcai/CornazGS13}.

\subsection{Hardness Results}
\label{subsec:hardness}

We start with the complexity analysis of voter deletion single-peaked consistency.
In the reduction we are going to cascade two or more preference profiles. The following definition captures this operation.

\begin{definition}
Let $(C_1,\mathcal{P}_1)$ and $(C_2,\mathcal{P}_2)$ be two elections with $C_1\cap C_2=\emptyset$.
Furthermore, let $\mathcal{P}_1= (\vote{1}',\ldots,\vote{n}')$ and $\mathcal{P}_2= (\vote{1}'',\ldots,\vote{n}'')$.
We define $\mathcal{P}_1\circlesucc \mathcal{P}_2= (\vote{1},\ldots,\vote{n})$, where
for any $1\leq i\leq n$ the total order $\vote{i}$ is defined by
\[
 c \succ_i c' \text{ iff } (c,c' \in C_1 \text{ and } c \succ'_i c') \text{ or }
 (c,c' \in C_2 \text{ and } c \succ''_i c') \text{ or }
 (c \in C_1 \text{ and } c' \in C_2).
\]
\end{definition}

Note that $\mathcal{P}_1\circlesucc \mathcal{P}_2$ is always  a preference profile on $C_1\cup C_2$.

\begin{lemma}\label{lem:merge-profiles-spc}
Let $(C_1,\mathcal{P}_1)$ and $(C_2,\mathcal{P}_2)$ be two elections with $C_1\cap C_2=\emptyset$.
Assume that: 
\begin{itemize}
\item $\mathcal{P}_1$ and $\mathcal{P}_2$ are single-peaked consistent with respect to the axes $A_1$ and $A_2$, respectively.
\item The votes in $\mathcal{P}_2$ have at most $2$ distinct top-ranked candidates.
\item These (two) top-ranked candidates are adjacent on the axis $A_2$.
\end{itemize}  
Then $\mathcal{P}_1\circlesucc \mathcal{P}_2$ is single-peaked.
\end{lemma}

\begin{proof}
We are going to construct an axis $A$ in a way that $\mathcal{P}_1\circlesucc \mathcal{P}_2$ is single-peaked with respect to $A$.
First we split $A_2$ in two parts $A_2'$ and $A_2''$.
If $\mathcal{P}_2$ contains votes with two distinct top-ranked candidates (which have to be adjacent), we split $A_2$ in between these two candidates.
If all votes in $\mathcal{P}_2$ share the same top-ranked candidate, we split $A_2$ left of this candidate (this is arbitrary).
The new axis $A$ is $A_2'$ followed by $A_1$ and then $A_2''$, i.e., $A_2'>A_1>A_2''$.
The correctness proof of this construction is straightforward.
\end{proof}

Before we start with the hardness proof, let us first make the following observation.

\begin{observation}\label{lem:profiles-spc}
  We are given a set of candidates $C=\{a,b,c,d\}$ and the following three votes:
$\vote{d}:acbd$, 
$\vote{e}:cbda$, and 
$\vote{ne}:dcba$.
  Then the preference profile $(\vote{d}, \vote{e})$ is single-peaked with respect to the axis $a>c>b>d$ and $(\vote{e}, \vote{ne})$ is single-peaked with respect to the axis $d>c>b>a$. The profile $(\vote{d}, \vote{ne})$ is not single-peaked consistent\footnote{Indeed, $(\vote{d}, \vote{ne})$ corresponds to an $\alpha$-configuration in \citeS{bal-hae:j:characterization-single-peaked} characterization of the single-peaked domain and thus cannot be single-peaked consistent.}.
\end{observation}

We now show $\np$-hardness via a reduction from the clique problem, a well-known $\np$-complete problem. %
We remark that the following result has been proven independently by \citeA{bre-che-woe-profiles-nearby} in a more general form that also applies to domain restrictions other than single-peakedness.

\EP{Clique}
{A graph $(V_G,E_G)$ and a positive integer $s$.}
{Does $(V_G,E_G)$ contain a clique of size $s$, i.e., has the graph $(V_G,E_G)$ an induced subgraph of size $s$ that is complete?}

\begin{theorem}\label{thm:k-voter-deletion-NPc}
\textsc{Voter Deletion Single-Peaked Consistency} is $\np$-complete.
\end{theorem}
\begin{proof}
To show hardness we reduce from \textsc{Clique}.
Let $V_G=\{v_1,\ldots,v_n\}$.
Each vertex $v_i$ has four corresponding candidates $c^1_i,\ldots,c^4_i$.
We consequently have $C=\{c^1_1,\ldots,c^4_1,c^1_2,\ldots,$ $c^4_2, \ldots,c^1_n,\ldots,c^4_n\}$.
The votes directly correspond to vertices and thus $\mathcal{P}=(\vote{1},\ldots,\vote{n})$.
In order to define the votes we introduce three functions creating partial votes.
For $a,b,c,d\in C$, let $f_d(a,b,c,d) = a c b d$, $f_{e}(a,b,c,d) = c b d a$, and $f_{ne}(a,b,c,d) = d c b a$.
If we consider $f_d$, $f_{e}$ and $f_{ne}$ as votes then by Observation~\ref{lem:profiles-spc} $(f_d, f_{e})$ and $(f_{e}, f_{ne})$ are single-peaked consistent but $(f_d, f_{ne})$ is not.
Next we define a function $p(i,j)$, mapping a pair in $\{1,\ldots,n\}^2$ to a total order on $\{c^1_j,\ldots,c^4_j\}$.

\[p(i,j)=\begin{cases}
  f_d(c_j^1,c_j^2,c_j^3,c_j^4) & \text{if $i=j$,}\\
  f_{e}(c_j^1,c_j^2,c_j^3,c_j^4) & \text{if $\{i,j\}\in E_G$,}\\
  f_{ne}(c_j^1,c_j^2,c_j^3,c_j^4) & \text{if $\{i,j\}\notin E_G$.}
  \end{cases}\]

The intuition behind function $p(i,j)$ is to encode a row of the adjacency matrix of $G$ as a vote in the preference profile $\mathcal{P}$.
To this end, we put in ``cell'' $(i,j)$ the result of $f_e$ if there is an edge between $i$ and $j$.
If there is no edge between $i$ and $j$, then we put the result of $f_{ne}$ in cell $(i,j)$.
In the special case $i=j$ (we are in the diagonal of the matrix) we put the result of $f_d$ in the cell.

Let the partial profiles representing the columns of the adjacency matrix be defined as $\mathcal{P}_j= (p(1,j),\ldots,p(n,j))$, for $1\leq j\leq n$.
We are now going to define the preference profile $\mathcal{P} = (\vote{1},\ldots,\vote{n})$ by $\mathcal{P} = \mathcal{P}_1 \circlesucc \mathcal{P}_2 \circlesucc \cdots \circlesucc \mathcal{P}_n$.

To conclude the construction let $E=(C,\mathcal{P})$ and $k= n-s$, i.e., we are allowed to delete $k$ voters from $E$ in order to obtain a single-peaked profile.
The intention behind the construction is that the voters in a single-peaked profile will correspond to a clique.
We claim that $G$ has a clique of cardinality $s$ if and only if it is possible to remove at most $k$ voters from $\mathcal{P}$ in order to make the resulting preference profile single-peaked consistent.

``$\Rightarrow$'' Assume that there is a clique $I=\{\vote{i_1},\ldots,\vote{i_s}\}$ with $|I|=s$.
Let $\mathcal{P}'=(\vote{i_1},\ldots,\vote{i_s})$.
Thereby we keep only those voters 
whose corresponding vertices are contained in the clique $I$.
Observe that the election $E'=(C,\mathcal{P}')$ can be obtained by deleting $k=n-s$ voters from the profile $\mathcal{P}$.
It remains to show that $\mathcal{P}'$ is indeed single-peaked consistent.
Let $C_i=\{c_i^1,c_i^2,c_i^3,c_i^4\}$ for $i\in\{1,\ldots,n\}$.
Since $I$ is a clique, for each ${\vote{x},\vote{y}}\in I$, $x\neq y$, there is an edge $\{x,y\}\in E_G$.
Thus, for every $i\in\{1,\ldots,n\}$, $\mathcal{P}'[C_i]$ either contains only instantiations of $f_d$ and of $f_{e}$ (if $c_i$ is in the clique) or only contains $f_e$ and $f_{ne}$ (if $c_i$ is not in the clique).
By Observation~\ref{lem:profiles-spc}, we conclude that $\mathcal{P}'[C_i]$ is single-peaked consistent.
Now we intend to use Lemma~\ref{lem:merge-profiles-spc} to show that also $\mathcal{P}'$ is single-peaked.
Note that $\mathcal{P}'[C_i]$ contains at most two distinct top-ranked candidates.
In addition, these two top-ranked candidates are adjacent on the axis which gives single-peaked consistency, as can be seen as follows:
Consider again Observation~\ref{lem:profiles-spc}.
For $(f_d,f_{e})$ the top-ranked candidates $a$ and $c$ are adjacent on the axis $a>c>b>d$.
The same holds for $(f_{e},f_{ne})$ with axis $d>c>b>a$ and $c$, $d$ as top-ranked candidates.
Since all conditions of Lemma~\ref{lem:merge-profiles-spc} are fulfilled, we can apply it iteratively.
Therefore, $\mathcal{P}'[C_1]\circlesucc\mathcal{P}'[C_2], (\mathcal{P}'[C_1]\circlesucc\mathcal{P}'[C_2])\circlesucc\mathcal{P}'[C_3],$ etc.\ are single-peaked consistent and hence also $\mathcal{P}'$ is single-peaked consistent.

``$\Leftarrow$'' Assume that
$E^*=(C,\mathcal{P}^*)$ is an election that has been obtained from $E$ by deleting at most $k$ voters from $\mathcal{P}$ such that $\mathcal{P}^*$ is single-peaked.
Then there exists also an election $E'=(C,\mathcal{P}')$ which is obtained from $E$ by deleting exactly $k$ voters from $\mathcal{P}$ such that $\mathcal{P}'$ is single-peaked.
This is because deleting additional voters from a single-peaked profile can never break single-peakedness.
Consequently $\mathcal{P}'$ contains $s$ votes.
Let $i_1,\ldots,i_s\in\{1,\ldots,n\}$ such that $\mathcal{P}'=(\vote{i_1},\ldots,\vote{i_s})$.
We claim that the vertices $\{v_{i_1},\ldots,v_{i_s}\}$ form a clique in $G$.
As before, let $C_i=\{c_i^1,c_i^2,c_i^3,c_i^4\}$ for $i\in\{1,\ldots,n\}$.
By Lemma~\ref{lem:spc-subsets} we know that for all $i\in\{1,\ldots,n\}$, $\mathcal{P}'[C_i]$ is single-peaked consistent.
Then, by Observation~\ref{lem:profiles-spc}, each column must not contain an instance of $f_d$ together with an instance of $f_{ne}$.
Let $j\in\{i_1,\ldots,i_s\}$.
Observe that by construction vote $\vote{j}$ contains an instance of $f_d$ on candidate set $C_j$.
Consequently, all other votes in $\mathcal{P}'$ have to be instantiations of $f_e$ on $C_j$ and thus vertex $v_j$ is adjacent to all other vertices in $\{v_{i_1},\ldots,v_{i_s}\}$.
Since $j$ is arbitrary, the vertices $v_{i_1},\ldots,v_{i_s}$ form a clique.
\end{proof}

We now turn to additional axes single-peaked consistency. Here we make use of a similar construction as presented in Theorem~\ref{thm:k-voter-deletion-NPc} with the difference that we now show $\np$-hardness via a reduction from the partition into cliques problem \cite{karp1972reducibility,gar-joh:b:int}.

\EP{Partition Into Cliques}
{A graph $(V_G,E_G)$ and a positive integer $s$.}
{Is it possible to partition $V_G$ into $s$ sets such that each set of vertices induces a clique on $(V_G,E_G)$?}

\begin{theorem}
\label{thm:k-AA-NPc}
\textsc{Additional Axes Single-Peaked Consistency} is $\np$-complete.
\end{theorem}
\begin{proof}
  Hardness is shown by a reduction from \textsc{Partition Into Cliques}.
  For the reduction we use the same transformation as presented in the proof of Theorem~\ref{thm:k-voter-deletion-NPc} to obtain an election.
  Then we set $k= s-1$, i.e., we are searching for a partition of the voters into $s$ disjoint sets such that each of the partitions is single-peaked consistent.
  Due to the one-to-one correspondence between voters and vertices we can use the partition of the vertices to obtain a partition of the voters and vice versa.
  With arguments similar to the proof of Theorem~\ref{thm:k-voter-deletion-NPc} one can show that a set of vertices is a clique if and only if the corresponding profile is single-peaked consistent.
\end{proof}

\begin{remark}
The \textsc{Partition Into Cliques} problem is \np-complete even when one is asked to partition the graph into three cliques.
Consequently it follows from the proof of Theorem~\ref{thm:k-AA-NPc} that \textsc{Additional Axes Single-Peaked Consistency} is $\np$-complete even for $k=2$, i.e., for deciding single-peaked consistency with two additional axes.
\end{remark}

In the proofs of our next two results, we will provide reductions from the $\np$-complete \textsc{Minimum Radius} problem~\cite{fra-lit:j:codes}.
It is defined as follows:

\EP{Minimum Radius}
{A set of strings $S\subseteq\{0,1\}^\ell$ and a positive integer $s$.}
{Has $S$ a radius of at most $s$, i.e., is there a string $\alpha\in\{0,1\}^l$ such that each string in $S$ has a Hamming distance to $\alpha$ of at most $s$?}

\begin{theorem}
\textsc{Local Candidate Deletion Single-Peaked Consistency} is $\np$-complete.
\label{thm:k-LCD-NPc}
\end{theorem}

\begin{proof}
For proving hardness, we reduce from the \textsc{Minimum Radius} problem.
Let $S\subseteq\{0,1\}^\ell$ and $s$ a positive integer.
Given a string $\beta\in S$, let $\beta(i)$ denote the bit value at the $i$-th position in $\beta$.
We are going to construct an \textsc{\dlcd Single-Peaked Consistency} instance.
Each string in $S=\{\beta_1,\ldots,\beta_n\}$ will correspond to two voters.
Each bit of the strings will correspond to two candidates.
In addition, we have $2\ell s+2$ extra candidates.
Consequently, we have $C=\{c^1_1,c^2_1,c^1_2,c^2_2,\ldots$, $c^1_\ell,c^2_\ell$, $c'_1,\ldots,c'_{\ell s+1}, c''_1,\ldots,c''_{\ell s+1}\}$.

We define the preference profile with the help of two functions creating total orders.
\begin{align*}
& f_{0}(a,b)= a b &
f_{1}(a,b)= b a
\end{align*}
The vote $\vote{j}$, for each $j\in \{1,\ldots,n\}$, is of the form
\[\vote{j}:c'_1\ \ldots\ c'_{\ell s+1}\ \ f_{\beta_j(1)}(c_1^1,c_1^2)\ \ f_{\beta_j(2)}(c_2^1,c_2^2)\ \ \ldots\ \  f_{\beta_j(\ell)}(c_\ell^1,c_\ell^2)\ \ c''_1\ \ldots\ c''_{\ell s+1}.\]
The preference profile $\mathcal{P}$ is now defined as 
$(\vote{1},\ldots,\vote{n},\overline{\vote{1}},\ldots,\overline{\vote{n}})$.
We claim that $(C,\mathcal{P})$ is $s$-local candidate deletion single-peaked consistent if and only if $S$ has a radius of at most $s$. 

``$\Leftarrow$'' Suppose that $S$ has a radius of at most $s$, i.e., there is a string $\alpha\in\{0,1\}^\ell$ with Hamming distance at most $s$ to each $\beta\in S$.
We consider the following axis $A$:
\[c'_1>\cdots>c'_{\ell s+1}>f_{\alpha(1)}(c_1^1,c_1^2)>f_{\alpha(2)}(c_2^1,c_2^2)>\cdots >f_{\alpha(\ell)}(c_\ell^1,c_\ell^2)>c''_1>\cdots>c''_{\ell s+1}.\]
We claim that $\mathcal{P}$ is single-peaked with respect to $A$ after deleting at most $s$ candidates in each vote.
The deletions for vote $\vote{j}$, $j\in\{1,\ldots,n\}$, are the following:
We delete candidate $c_i^1$ in $\vote{j}$ if and only if $\alpha(i)\neq \beta_j(i)$.
The deletions in 
$\overline{\vote{j}}$ 
are exactly the same as in $\vote{j}$.
These are at most $s$ deletions since the Hamming distance between $\alpha$ and every $\beta\in S$ is at most $s$.
After these deletions all votes are either subsequences of $A$ or its reverse.
Hence we obtain a single-peaked consistent profile.

``$\Rightarrow$'' Let $\mathcal{P'}$ be the partial, single-peaked consistent profile that was obtained by deleting at most $s$ candidates in each vote.
First, note that some $c'\in\{c'_1,\ldots,c'_{\ell s+1}\}$ has not been
deleted in any vote since in total at most $\ell\cdot s$ different candidates can be deleted.
In the same way let $c''\in\{c''_1,\ldots,c''_{\ell s+1}\}$ be a candidate that has not been deleted in any vote.
Now let us consider the profile $\mathcal{P'}[\{c',c'',c_i^1,c_i^2\}]$ for any $i\in\{1,\ldots,\ell\}$.
We claim that $\alpha\in\{0,1\}^\ell$, defined in the following way, has a Hamming distance of at most $s$ to all bitstrings in $S$. 
\[\alpha(i)=\begin{cases} 0 &\mbox{if } \mathcal{P'}\text{ contains a vote with } c' c_i^1 c_i^2 c'',\\
1 &\mbox{if } \mathcal{P'}\text{ contains a vote with } c' c_i^2 c_i^1 c'',\\
1 &\mbox{otherwise.} \end{cases}\]
First, observe that Case 1 and 2 cannot occur at the same time since then $\mathcal{P'}$ would not be single-peaked consistent.

Let $\beta_j\in S$, $j\in\{1,\ldots,n\}$.
Note that if at any position $i$, $\beta_j(i)\neq\alpha(i)$ then either $c_i^1$ or $c_i^2$ had to be deleted in the vote $\vote{j}$.
Otherwise $\mathcal{P}'$ would not be single-peaked consistent.
Hence $\card{\{i\in\{1,\ldots,\ell\}\mid \alpha(i)\neq \beta_j(i)\}}\leq s$ because otherwise we would require more than $s$ candidate deletions in the corresponding vote $\vote{j}$.
Hereby we have shown that the Hamming distance of $\alpha$ and $\beta_j$ is at most $s$.
\end{proof}

\begin{theorem}\label{thm:k-LS-NPc}
\textsc{Local Swaps Single-Peaked Consistency} is \np-complete.
\end{theorem}

\begin{proof}
We use the same construction as in the proof of Theorem~\ref{thm:k-LCD-NPc}.
It holds that $(C,\mathcal{P})$ is $s$-local swaps single-peaked consistent if and only if $S$ has a radius of at most $s$. 
This can be shown similarly to the proof of Theorem~\ref{thm:k-LCD-NPc} except that we swap candidates instead of deleting them.
\end{proof}

The following problem will be useful for showing \np-hardness of \textsc{Global Swaps Single-Peaked Consistency}.
Given two votes, $\vote{x}$ and $\vote{y}$, let $\mathsf{swaps}(\vote{x},\vote{y})$ denote the minimum number of swaps of adjacent candidates needed to make $\vote{x}$ and $\vote{y}$ equal, i.e., $\mathsf{swaps}(\vote{x},\vote{y})$ is the Kendall tau distance of $\vote{x}$ and $\vote{y}$.

\EP{Kemeny Optimal Aggregation}
{An election $(C,\mathcal{P})$, with $\mathcal{P}=(\vote{1},\ldots,\vote{n})$, and an integer $s$.}
{Is there a vote $\vote{}^*$ on $C$ such that $\sum_{1\leq i\leq n}\mathsf{swaps}(\vote{i},\vote{}^*) \leq s$.}

\textsc{Kemeny Optimal Aggregation} was shown to be \np-complete \cite{bar-tov-tri:j:voting-schemes}.
Later, this result was strengthened to require only four voters \cite{dwo-kum-nao-siv:c:rank-aggregation,biedl2009complexity}.

\begin{theorem}\label{thm:k-GS-NPc}
\textsc{Global Swaps Single-Peaked Consistency} is \np-complete.
\end{theorem}
\begin{proof}
  We show \np-hardness of this problem by reduction from \textsc{Kemeny Optimal Aggregation}.
  Let an instance of \textsc{Kemeny Optimal Aggregation} be given by $C=\{c_1,\ldots,c_m\}$, $\mathcal{P}=(\vote{1},\ldots,\vote{n})$, and $s$.
  We define $k$, the number of allowed swaps, be defined as $2s$.
  Then we create a new election $(C',\mathcal{P}')$ with $C'=C\cup\{c_1^\text{top},\ldots,c_{2k+1}^\text{top},$ $c_1^\text{last},\ldots,c_{2k+1}^\text{last}\}$, i.e., $\card{C'}=m+4k+2$.
  For each $i\in\{1,\ldots,m\}$ we create two votes $\vote{i}'$ and $\overline{\vote{i}'}$ as follows.
  The vote $\vote{i}'$ ranks $c_1^\text{top}$ first, followed by $c_2^\text{top}$, $c_3^\text{top}$, $\ldots$, $c_{2k}^\text{top}$ and finally $c_{2k+1}^\text{top}$.
  Then it ranks the candidates in $C$ in the same order as $\vote{i}$ does.
  Finally, it orders the candidates $c_1^\text{last}\ldots c_{2k+1}^\text{last}$ with descending preference, i.e., $c_{2k+1}^\text{last}$ being the last-ranked candidate.
  Vote $\overline{\vote{i}'}$ is the reverse of $\vote{i}'$.
  The preference profile $\mathcal{P}'$ is now defined as $(\vote{1}', \overline{\vote{1}'},\ldots,\vote{n}',\overline{\vote{n}'})$.
  We refer to $\vote{1}', \ldots,\vote{n}'$ as the non-reversed votes and to $\overline{\vote{1}'},\ldots,\overline{\vote{n}'}$ as the reversed votes.
  We claim that $(C',\mathcal{P}')$ is $k$-global swaps single-peaked consistent if and only if $(C,\mathcal{P})$ and $s$ are a yes-instance of the \textsc{Kemeny Optimal Aggregation} problem.

  ``$\Rightarrow$'' Suppose that $(C',\mathcal{P}')$ is $k$-global swaps single-peaked consistent.
  Therefore, one can obtain a profile $\mathcal{P}^S$ from $\mathcal{P}'$ by applying at most $k=2s$ swaps such that $\mathcal{P}^S$ is single-peaked consistent with respect to an axis $A$.
  Since there are $2k+1$ candidates in the set $\{c_1^\text{top},\ldots,c_{2k+1}^\text{top}\}$ at least one of them must have remained in place in each vote.
  Analogously, the same holds for one of the candidates contained in the set $\{c_1^\text{last},\ldots,c_{2k+1}^\text{last}\}$.
  Let $c_\text{top}$ and $c_\text{last}$ denote these two candidates.
  From Lemma~\ref{lem:spc-subsets} we know that $\mathcal{P}^S[\{c_\text{top},c_1,\ldots,c_m,c_\text{last}\}]$ is single-peaked consistent as well.
  Observe that all non-reversed votes in $\mathcal{P}^S[\{c_\text{top},c_1, \ldots,$ $c_m,c_\text{last}\}]$ have $c_\text{top}$ as top-ranked candidate and $c_\text{last}$ as last candidate, while in all reversed votes $c_\text{last}$ is top and $c_\text{top}$ is the last-ranked candidate.
  By Lemma~\ref{lem:1..n,n..1} all non-reversed votes in $\mathcal{P}^S[\{c_1,\ldots,c_m\}]$ must be ordered in the same way and the reversed votes in exactly their reverse order.
  We denote this ordering of $\{c_1,\ldots,c_m\}$ by $\vote{}^*$.
  Notice that turning the non-reversed votes into $\vote{}^*$ requires the same number of swaps as turning the reversed votes into $\overline{\vote{}^*}$.
  Therefore, $\frac{k}{2}=s$ swaps are sufficient to turn all non-reversed votes into $\succ^*$.
  Taken together, $\vote{}^*$ fulfills all properties to be a yes-instance of the \textsc{Kemeny Optimal Aggregation} problem.

  ``$\Leftarrow$'' Assume $(C,\mathcal{P})$ and $s$ describe a yes-instance of the \textsc{Kemeny Optimal Aggregation} problem.
  Then there is some common ordering $\vote{}^*$, which has in total a swap distance of $\leq s$ to all votes in $\mathcal{P}$.
  Then, $(C',\mathcal{P}')$ is $k$-global swaps single-peaked consistent with respect to the axis $c_1^\text{top}>c_2^\text{top}>\cdots>c_{2k+1}^\text{top}>[\text{$c_1,\ldots, c_n$ as ordered by $\vote{}^*$}]>c_1^\text{last}>c_2^\text{last}>\cdots>c_{2k+1}^\text{last}$.
  This is because all votes can be brought into the form $c_1^\text{top}\succ c_2^\text{top}\succ\cdots\succ c_{2k+1}^\text{top}\succ[\text{$c_1,\ldots, c_n$ as ordered by $\vote{}^*$}]\succ c_1^\text{last}\succ c_2^\text{last}\succ\cdots\succ c_{2k+1}^\text{last}$ or its reverse by using at most $k=2s$ swaps -- $s$ swaps for the non-reversed and $s$ swaps for the reversed votes.
\end{proof}

\begin{remark}
Since \textsc{Kemeny Optimal Aggregation} with only four voters is \np-complete \cite{dwo-kum-nao-siv:c:rank-aggregation}, it follows from the proof of Theorem~\ref{thm:k-GS-NPc} that \textsc{Global Swaps Single-Peaked Consistency} is \np-complete even for eight voters.
\end{remark}

\subsection{A Polynomial-Time Algorithm for Candidate Deletion Single-Peaked Consistency}
\label{subsec:cd-poly-alg}

In contrast to the previous hardness results, we are able to show that \textsc{Candidate Deletion Single-Peaked Consistency} can be decided in polynomial time.
The algorithm builds upon the $\mathcal{O}(n\cdot m)$ time algorithm for testing single-peaked consistency by Escoffier, Lang and \"Ozt\"urk~\citeyear{esc-lan-oez:c:single-peak}.
Since we make some modifications to the algorithm and also for the sake of completeness we present it here as well.

For the remainder of this section let $(C,\mathcal{P})$ be an election with $n$ voters and $C=\{c_1,\ldots,c_m\}$.

\subsubsection{The Single-Peaked Consistency Algorithm}
The algorithm presented here is a modified version of the algorithm by \citeA{esc-lan-oez:c:single-peak}. It tests whether a given profile is single-peaked consistent and thus does not consider candidate deletions.
Before we start with presenting the algorithm, let us fix some notation.

\begin{definition}
For $C'\subseteq C$, let $L(\mathcal{P},C')$ denote the set of last-ranked candidates in $\mathcal{P}[C']$.
\label{def:lastranked}
\end{definition}

\begin{definition}
An \emph{incomplete axis} is a total order on a subset of $C$ with a marked position that indicates where further elements may be added.
We denote this position by a star symbol, e.g., the incomplete axis $c_1>c_2>\star>c_3$ allows additional candidates to be added right of $c_2$ and left of $c_3$.
We write $|A|$ to denote the number of candidates on an incomplete axis $A$.
The \emph{boundary} of an incomplete axis $A$, $\textsf{boundary}(A)$, is a quadruple consisting of the two candidates left of the star and the two candidates right of the star, e.g., $\textsf{boundary}(c_1>c_2>\star>c_3>c_4>c_5)=(c_1,c_2,c_3,c_4)$.
If only one or no candidates exist left/right of the star, the corresponding entries in the quadruple are $\epsilon$, e.g., $\textsf{boundary}(c_1>\star)=(\epsilon,c_1,\epsilon,\epsilon)$.

Given an incomplete axis $A$ and a candidate set $C$, an axis \emph{$A'$ extends $A$} if $A'$ can be constructed from $A'$ by adding elements left or right of the $\star$ symbol.
\end{definition}

The algorithm proceeds iteratively by placing the last-ranked candidates that have not yet been placed.
Let $C'\subseteq C$ be the set of candidates that have not yet been positioned on the (incomplete) axis $A$.
The algorithm checks what kind of constraints follow from each vote.
If these constraints do not contradict each other, the set of last-ranked candidates $L(\mathcal{P},C')$ is placed.
We denote this procedure by $\textsf{place}(A,X)$ where $X=L(\mathcal{P},C')$.
The procedure $\textsf{place}(A,X)$ returns either a new incomplete axis (extending $A$ by the candidates in $X$) or the value $\texttt{inconsistent}$.
The algorithm repeatedly invokes $\textsf{place}$ until all elements have been placed or a contradiction has been found.

Now we describe $\textsf{place}(A,X)$ in detail since it is also used by our candidate deletion algorithm.
Let $\textsf{boundary}(A)=(b_1',b_1,b_2,b_2')$, i.e., the current incomplete axis $A$ is given as $\cdots>b_1'>b_1> \star > b_2>b_2'>\cdots$.
If a condition contains a boundary element and this element is $\epsilon$ (i.e., it does not exist), corresponding constraints can be ignored.
The following cases are considered for each vote $\succ_i$, $i\in\{1,\ldots,n\}$.
\begin{enumerate}[{Case} 1:]
\item $\card{L(\mathcal{P},C')}\geq 3$. 
There are three or more candidates that would have to be placed at the positions next to $b_1$ and $b_2$.
Since this is not possible, $\mathcal{P}$ is not single-peaked consistent; \textsf{place} returns \texttt{inconsistent}.
\item $L(\mathcal{P},C') = \{x_1,x_2\}$.
The candidates $x_1$ and $x_2$ have to be placed at the positions next to $b_1$ and next to~$b_2$.
	\begin{enumerate}
	\item $b_1 \succ_i x_1$ and $b_2 \succ_i x_1 $: 
	This case cannot occur since $x_1$ is ranked below $b_1$ and $b_2$ and thus cannot be placed after $b_1$ and $b_2$.
	\item $x_1 \succ_i b_1$ and $x_1 \succ_i b_2 $: 
	There are no constraints for $x_1$ that follow from $\vote{i}$.
	\item $b_2 \succ_i x_1$ and $x_2\succ_i x_1$: 
	In this case $x_1$ has to be placed next to $b_1$ and therefore $x_2$ is placed next to $b_2$.
	\item $b_1 \succ_i x_1$ and $x_2\succ_i x_1$: 
	In this case $x_1$ has to be placed next to $b_2$ and therefore $x_2$ is placed next to $b_1$.
	\end{enumerate}
All these rules are also applicable if $x_1$ and $x_2$ are interchanged.

\item $L(\mathcal{P},C') = \{x\}$.
The candidate $x$ has to be placed either at the position next $b_1$ or $b_2$. It is important that if $x$ is the last candidate to be placed, it can be placed either next to $b_1$ or next to $b_2$.
	\begin{enumerate}
	\item $b_1 \succ_i x$ and $b_2 \succ_i x$: 
	This case cannot occur since $x$ is ranked below $b_1$ and $b_2$ and thus cannot be placed after $b_1$ and $b_2$.
	\item $x \succ_i b_1$ and $x \succ_i b_2$: 
	There are no constraints for $x$.
	\item $b_2 \succ_i x $: 
	In this case $x$ has to be placed left, i.e., next to $b_1$.
	\item $b_1 \succ_i x $: 
	Then $x$ has to be placed right, i.e., next to $b_2$.
	\end{enumerate}
\end{enumerate}
In addition to these three cases, the following constraint is applicable independently of the cardinality of $L(\mathcal{P},C')$.
Let $x\in L(\mathcal{P},C')$.
\begin{enumerate}[{Case 4}:]
\item If $b_1'\succ_i b_1$ and $x\succ_i b_1$, or if $b_2'\succ_i b_2$ and $x\succ_i b_2$, then the candidates $b_1',b_1,x$ or $x, b_2,b_2'$, respectively, violate the single-peaked condition (cf.\ Lemma~\ref{lem:not-spc-crit}). Thus \textsf{place} returns $\texttt{inconsistent}$.
\end{enumerate}

For each vote $\succ_i$, these case distinctions yield constraints on placing the candidates in $X$.
If there is a way to place the candidates in $X$ that is compatible with every vote, $\textsf{place}(A,X)$ has been successful and returns the new incomplete axis.
(If there is more than one possibility to place the candidates in $X$, $\textsf{place}$ chooses arbitrarily.)
Otherwise the value $\texttt{inconsistent}$ is returned.
To simplify the notation, we define $\textsf{place}(A,\emptyset)$ to return $A$.

As mentioned before, the $\textsf{place}$ procedure described here is similar to the procedure used in the original single-peaked consistency algorithm by \citeA{esc-lan-oez:c:single-peak}. The main difference is that the original single-peaked consistency algorithm places all remaining candidates at once as soon there is only one possibility left; $\textsf{place}$ continues to place at most two candidates in each step.
In particular, this necessitates Case 4, which did not appear in the original algorithm.

The following lemma shows that the $\textsf{place}$ procedure is indeed correct; the proof can be found in the appendix.

\begin{restatable}{lemma}{lemplacecorrect}\label{lem:place-correct}
Let $(C,\mathcal{P})$ be an election.
If the iterative application of $\textsf{place}$ yields an axis $A$, then $(C,\mathcal{P})$ is single-peaked with respect to $A$.
If at some point $\textsf{place}$ returns $\texttt{inconsistent}$, then $(C,\mathcal{P})$ is not single-peaked.
\end{restatable}

The following observation is the main reason why we can employ dynamic programming in our algorithm for deciding the \textsc{Candidate Deletion Single-Peaked Consistency} problem. 

\begin{observation}
The \textsf{place}$(A,X)$ procedure places the candidates in $X$ on the incomplete axis $A$ only by considering \textsf{boundary}$(A)$ and thus \textsf{place} does not depend on the full incomplete axis $A$.
\label{obs:boundary}
\end{observation}

\subsubsection{The Candidate Deletion Algorithm}
Observation~\ref{obs:boundary} states that the (at most) four boundary candidates of an incomplete axis fully determine whether and which further candidates can be placed on the axis.
The main idea of our algorithm is to store only incomplete axes that differ in these four candidates, i.e., only incomplete axes with differing boundaries.
If two axes with the same boundary are considered, we take the axis with the larger cardinality, i.e., the one with more candidates placed on it.
This strategy allows us to use a dynamic programming approach.

Our algorithm resembles the previously described single-peaked consistency algorithm in that it places last-ranked candidates first.
However, since we are allowed to delete candidates, our algorithm does not terminate if at some point three or more last-ranked candidates are encountered (cf.\ Case 1 in the single-peaked consistency algorithm).
Nevertheless, our algorithm utilizes the \textsf{place} procedure and thus can place at most two candidates in each step (cf.\ Lemma~\ref{lem:three-last-ranked}).
Candidates that would violate the single-peaked condition are not actively deleted but rather cannot be successfully placed.

We now describe the algorithm in more detail (cf.\  Algorithm~\ref{alg:kCD-SPC}).
The algorithm's main data structure is an associative array $\mathcal{S}$ containing incomplete axes.
The axes are indexed by quintuples $(c_1,c_2,c_3,c_4,Y)$ where $c_1,c_2,c_3,c_4\in C$ and $Y\subseteq \{c_2,c_3\}$, i.e., such quintuples constitute keys for $\mathcal{S}$.
The associative array my contain at index $(c_1,c_2,c_3,c_4,Y)$ only an incomplete axis $A$ with $\textsf{boundary}(A')=(c_1,c_2,c_3,c_4)$.
Furthermore, the candidates in $Y$ have been placed most recently on $A$; this can either be $\{c_2\}$, $\{c_3\}$ or $\{c_2,c_3\}$.
We write $\mathcal{S}[c_1,c_2,c_3,c_4,Y]$ to denote the entry $\mathcal{S}$ with key $(c_1,c_2,c_3,c_4,Y)$.

{\renewcommand{\baselinestretch}{1.1}
\SetKwFor{OwnRepeat}{repeat}{times}{end}
\begin{algorithm}[t]
  \DontPrintSemicolon
  $\mathcal{S}_0[\epsilon,\epsilon,\epsilon,\epsilon,\emptyset]\leftarrow \star$\tcp*[r]{$\mathcal{S}_0$ contains the empty incomplete axis}
  \For{$i = 1\ldots m$}{
	$\mathcal{S}_i\leftarrow\mathcal{S}_{i-1}$\;
    \ForEach{incomplete axis $A\in\mathcal{S}_{i-1}$} {
      Let $(c_1,c_2,c_3,c_4,Y)$ be the key of $A$.\tcp*[r]{i.e., $\mathcal{S}_{i-1}(c_1,c_2,c_3,c_4,Y)=A$}
      \ForEach{$X \in N(i, Y)$} {
        	$A_\mathit{new}\leftarrow \textsf{place}(A,X)$\;
        \If{\textnormal{$A_\mathit{new}\neq\texttt{inconsistent}$}} {
			\eIf{$\mathcal{S}_{i}[c_1,c_2,c_3,c_4,X]$ is empty}
			 {$\mathcal{S}_{i}[c_1,c_2,c_3,c_4,X]\leftarrow A_\text{new}$}
			 {\If{$\card{A_\text{new}} > \card{\mathcal{S}_{i}[c_1,c_2,c_3,c_4,X]}$\nllabel{alg:choice1}}
			   {$\mathcal{S}_{i}[c_1,c_2,c_3,c_4,X]\leftarrow A_\text{new}$\nllabel{alg:choice2}}
			 }
        }
      }
    }
  }
  \Return{an axis $A$ in $\mathcal{S}_m$ with maximum $\card{A}$}
\caption{Polynomial-time algorithm for $k$-candidate deletion single-peaked consistency -- Theorem~\ref{thm:k-GD-poly}}
\label{alg:kCD-SPC}
\end{algorithm}
} %

We start with $\mathcal{S}_0$ containing only the empty incomplete axis $\star$.
(Recall that $\star$ marks the position where new candidates can be added to the axis.)
We write $N(i,Y)$ to denote the set containing all $X\subseteq C$ that may be placed on $A$ in step $i$ of the algorithm; we will define $N(i,Y)$ after an overview of the algorithm.
Note that at most two of the candidates can be placed and hence $1\leq |X|\leq 2$.
So, in step $i=1$, for every axis $A\in \mathcal{S}_0$ (only $\star$) and every set of candidates $X\in N(1,Y)$, we use the \textsf{place} procedure to place the candidates in $X$ on $A$.
This step gives rise to new incomplete axes.
These new axes as well as those in $\mathcal{S}_{0}$ are stored in $\mathcal{S}_{1}$.

For $i=2$, we continue by placing the candidates in $N(2,Y)$ for every incomplete axis $A\in\mathcal{S}_1$.
Again this creates new incomplete axes, which we store in $\mathcal{S}_{2}$ (in addition to elements from $\mathcal{S}_1$).
At this point, it might be that $\mathcal{S}_{2}$ already contains an entry with the same key.
In this case, we keep the axis with more candidates placed on it.
We repeat this procedure until we have considered the sets in $N(m,Y)$.
The associative array $\mathcal{S}_m$ now contains a several incomplete axes; we output a cardinality maximal axis as it requires the minimum number of candidate deletions.

\smallskip
Let us now define $N(i,Y)$.%
\begin{definition}\label{def:nia}
For all $1\leq i\leq m$, let $L_i = L\left(\mathcal{P},C\setminus( L_1\cup\cdots\cup L_{i-1})\right)$.
Further, fix $1\leq i\leq m$ and let $Y$ be the candidates last placed on the incomplete axis.
We define $N(i,Y)$ to be the set of all $X\subseteq L_i\cup\dots\cup L_m$ satisfying the following conditions:
\begin{multicols}{2}
\begin{enumerate}[(i)]
\item $1\leq |X| \leq 2$
\item $X\cap L_i\neq \emptyset$
\item $L(\mathcal{P},X)=X$
\item if $Y\neq\emptyset$ then $L(\mathcal{P},Y\cup X)= Y$
\end{enumerate}
\end{multicols}
\end{definition}
Note that $\bigcup_{i\in\{1,\ldots,m\}}L_i=C$ and that some $L_i$'s may be empty (more precisely, there might exist some index $j\leq m$ such that $L_j,\dots,L_m$ are empty).
The conditions can be intuitively understood as follows:
Condition (i) is a necessity if we want to obtain a single-peaked axis: No more than two last-ranked candidates may be placed on the axis.
Condition (ii) is mostly important for showing runtime bounds.
Note that condition (iii) implies that if $X=\{x_1,x_2\}$, then there must be a vote $\succ_i$ with $x_1\succ_i x_2$ and a vote $\succ_j$ with $x_2\succ_j x_1$. If one candidate would always be ranked below the other, the former had to be placed first and the second only in a later step. Condition (iv) guarantees that the candidates in $X$ are never ranked below $Y$. Candidates that are below $Y$ cannot be placed anymore, i.e., they have been deleted.

Intuitively, the correctness of Algorithm~\ref{alg:kCD-SPC} rests on two pillars. First, we have observed that the $\textsf{place}$ procedure only considers the four boundary elements of an incomplete axis. Hence, it is correct to identify incomplete axes with the same boundary. Since we search for a cardinality maximal axis, we can safely pick a larger axis if we have the choice (cf.\ Lines~\ref{alg:choice1} and \ref{alg:choice2}). Second, the correctness of the algorithm relies fundamentally on the correctness of the $\textsf{place}$ procedure, which we have shown in Lemma~\ref{lem:place-correct}.
We will now formally prove the runtime bounds and correctness of Algorithm~\ref{alg:kCD-SPC}.

\begin{theorem}\label{thm:k-GD-poly}
\textsc{Candidate Deletion Single-Peaked Consistency} can be solved in time $\mathcal{O}(n\cdot m^6)$.
\end{theorem}
\begin{proof}
The runtime bound can be seen as follows.
The sets $L_1,\ldots, L_{m}$ can be computed in $\mathcal{O}(m\cdot n)$ time.
The associative array $\mathcal{S}_i$ uses quintuples $(c_1,c_2,c_3,c_4,Y)$ as keys, where $Y\subseteq \{c_2,c_3\}$
Hence $\mathcal{S}_i$ contains at most $4 m^4$ entries.
Let us now bound the size of $N(i,Y)$.
The singleton sets in $N(i,Y)$ are subsets of $L_i$.
Two-element sets in $N(i,Y)$ contain one element of $L_i$ and one of $L_i\cup\dots\cup L_m$.
Thus, $|N(i,Y)|\leq |L_i|+ m|L_i|=(m+1)|L_i|$.
The \textsf{place} procedure is executed in step $i$ of the algorithm for every axis in $S_{i-1}$ and element of $N(i,Y)$, i.e., at most $4m^4  (m+1)|L_i|$ times.
If we add up over all iterations we obtain $\sum_{i=1}^m 4m^4 (m+1)|L_i| = 4m^4(m+1) \cdot \sum_{i=1}^m |L_i| = \mathcal{O}(m^6)$ procedure calls.
Since $\textsf{place}$ has a runtime of $\mathcal{O}(n)$, we require in total $\mathcal{O}(n\cdot m^6)$ time for  \textsf{place} calls.
Furthermore, $N(i,Y)$ has to be computed for every axis under consideration. It is straight-forward to verify that $N(i,Y)$ can be computed in $\mathcal{O}(n \cdot m\cdot |L_i|)$ time: At most $(m+1) |L_i|$ sets have to be considered and the conditions of Definition~\ref{def:nia} can be verified in $\mathcal{O}(n)$ time.
Again, adding up over all iterations we obtain a runtime of $\mathcal{O}(\sum_{i=1}^m m^4 nm|L_i|) = \mathcal{O}(n m^5 \cdot \sum_{i=1}^m |L_i|) = \mathcal{O}(nm^6)$, also within the desired time bound.

Let us now prove correctness.
For the one direction, let $A$ be a cardinality maximal axis returned by the algorithm.
Furthermore, let $X_1, \dots, X_m$ be the set of candidates placed in each step of the algorithm to obtain axis $A$.
In particular, this implies that $A$ is a total order on $X_1\cup \dots\cup X_m$.
Since some of those sets may be empty, let $X_1', \dots, X_\ell'$ be the selection of non-empty sets.
Let $\mathcal{P}'=\mathcal{P}[X_1'\cup \dots\cup X_m']$.
Our goal is to show that $\mathcal{P}'$ is single-peaked with respect to $A$ and we intend to apply Lemma~\ref{lem:place-correct} to prove this.

To do so, we will first show for every $i$, $1\leq i\leq \ell$, that $L(\mathcal{P}',X_i'\cup\dots\cup X'_\ell)=X'_i$.
We prove this statement by induction.
In the following we write $N(\cdot ,Y)$ and by that omit the first parameter, as the exact step at which a candidate was placed is irrelevant for our argument.
Observe that $X'_\ell\in N(\cdot ,X'_{\ell-1})$ since $X'_\ell$ has been placed on $A$ and $X'_{\ell-1}$ has been placed on $A$ before that. Thus the conditions listed in Definition~\ref{def:nia} apply to $X'_\ell$.
In particular $L(\mathcal{P}', X'_\ell)=X_\ell'$ holds due to condition (iii), which serves as our base case. %
For the induction step, assume that $L(\mathcal{P}',X'_{i+1}\cup\dots\cup X'_m)=X'_{i+1}$. We want to show that $L(\mathcal{P}',X'_{i}\cup\dots\cup X'_m)=X'_{i}$.
Note that for arbitrary $X', X''$ it holds that $L(\mathcal{P}',X'\cup X'')=L(\mathcal{P}',X'\cup L(\mathcal{P}',X''))$.
Thus, by our hypothesis, we know that $L(\mathcal{P},X'_{i}\cup\dots\cup X'_m)=L(\mathcal{P},X'_{i}\cup X'_{i+1})$.
Since $X_{i+1}'\in N(\cdot ,X'_i)$, condition (iv) yields that $L(\mathcal{P},X'_{i}\cup X'_{i+1})=X'_i$.

It follows now immediately from Lemma~\ref{lem:place-correct} that $\mathcal{P}'$ is single-peaked with respect to $A$:
We have shown that Algorithm~\ref{alg:kCD-SPC} first computes $\textsf{place}(\emptyset, X_1')$ with $X_1'=L(\mathcal{P}',X_1'\cup \dots\cup X_m')$. Then $\textsf{place}$ is applied on the resulting axis and $X_2'=L(\mathcal{P}',X_2'\cup \dots\cup X_m')$ is placed on it.
This procedure is repeated until we obtain axis $A$ and Lemma~\ref{lem:place-correct} applies. Thus, $(\mathcal{P}',X_1\cup \dots\cup X_m)$ is single-peaked with respect to $A$.

For the other direction, let $C'$ be a cardinality maximal subset of candidates such that $\mathcal{P}'=\mathcal{P}[C']$ is single-peaked.
Our goal is to show that $\mathcal{S}_m$ contains an axis with the same cardinality as $C'$.
Let $X_i = L\left(\mathcal{P}',C'\setminus( X_1\cup\cdots\cup X_{i-1})\right)$ for all $1\leq i\leq m$ and $X_0=\emptyset$.
Further let $\ell$ be the largest index such that $X_\ell\neq\emptyset$.
Let us consider $X_i$ and let $s(i)=\min \{1\leq j \leq m: x\in X_i\text{ and }x\in L_j\}$.
Our first goal is to show that $X_i\in N(s(i),X_{i-1})$.
To show this we have to check the conditions in Definition~\ref{def:nia}. First, $X_i\subseteq L_{s(i)}\cup\dots\cup L_m$ by definition of $s(i)$. Condition~(i) holds since $\mathcal{P}[C']$ is single-peaked (cf.~Lemma~\ref{lem:three-last-ranked}). Condition~(ii) holds also by choice of $s(i)$. Condition~(iii) holds because $X_i = L\left(\mathcal{P}',C'\setminus( X_1\cup\cdots\cup X_{i-1})\right)= L\left(\mathcal{P}', X_i\cup\cdots\cup X_{\ell}\right)$ and thus also $X_i = L\left(\mathcal{P}',X_i\right)$. Finally, condition~(iv) holds because $X_{i-1} = L\left(\mathcal{P}',C'\setminus( X_1\cup\cdots\cup X_{i-2})\right)=L\left(\mathcal{P}', X_{i-1}\cup\cdots\cup X_{\ell}\right)$ and thus also $X_{i-1} = L\left(\mathcal{P}', X_{i-1}\cup X_{i}\right)$.

Let us assume that the \textsf{place} procedure is applied to $\mathcal{P}'$. First $X_1$ is placed, then $X_2$, etc. Let $A_1, \dots, A_\ell$ be the corresponding incomplete axes.
Furthermore let $A_0$ be the empty incomplete axis, $X_0=\emptyset$ and $s(0)=0$.
We are going to show that for every $0\leq i\leq \ell$, there exists an incomplete axis $A'_i\in \mathcal{S}_{s(i)}$ such that $\textsf{boundary}(A_i')=\textsf{boundary}(A_i)$ and $|A_i'|\geq |A_i|$.
We prove this statement by induction.
For $i=0$, note that $A_0=\star\in \mathcal{S}_0$.
For the induction step, assume that $A'_i\in \mathcal{S}_{s(i)}$ such that $\textsf{boundary}(A_i)=\textsf{boundary}(A_i')$ and $|A_i|=|A_i'|$.
By our previous argument, it holds that $X_{i+1}\in N(s(i+1),X_i)$.
Since $A'_i\in \mathcal{S}_{s(i)}$, we compute $\textsf{place}(A_i',X_{i+1})$.
By Observation~\ref{obs:boundary}, $\textsf{place}(A_i',X_{i+1})$ is successful (i.e., it does not return \texttt{inconsistent}) if and only if $\textsf{place}(A_i,X_{i+1})$ is successful since both have the same boundary.
Let $A'_{i+1}$ be the axis returned by $\textsf{place}(A_i',X_{i+1})$.
Then $|A_{i+1}|=|A_{i+1}'|$ since in both cases the same candidates have been placed.
Also $\textsf{boundary}(A_{i+1})=\textsf{boundary}(A_{i+1}')$, again by Observation~\ref{obs:boundary}.
If $A_{i+1}'$ is stored in $\mathcal{S}_{i+1}$, we have proven the induction step.
If it is not stored, then there exists an $A^*_{i+1}$ in $\mathcal{S}_{i+1}$ with $\textsf{boundary}(A_{i+1})=\textsf{boundary}(A^*_{i+1})$ and $|A^*_{i+1}|\geq |A_{i+1}|$. Also in this case the induction statement holds.

We have shown that there exists an incomplete axis $A'_\ell\in \mathcal{S}_{s(\ell)}$ such that $\textsf{boundary}(A_\ell)=\textsf{boundary}(A_\ell')$ and $|A_\ell'|\geq |A_\ell|$.
If such an element exists in $\mathcal{S}_{s(\ell)}$, it does exist in $\mathcal{S}_m$.
Since $|A_\ell|=|C'|$, we have shown that the algorithm returns an axis of maximum cardinality.
\end{proof}

Finally, let us remark that very recently a modification of this algorithm has been proposed \cite{Prz16} that improves the runtime to $\mathcal{O}(n\cdot m^3)$.

\subsection{Complexity of Nearly Single-Peaked Evaluation}
\label{sec:complexity-eval}
In the previous sections we have analyzed the computational complexity of the \textsc{X Single-Peaked Consistency} problem.
We now turn to the computational complexity of the related \textsc{X Single-Peaked Evaluation} problem, where the axis is additionally given in the input.
Due to this additional information, all \textsc{X Single-Peaked Evaluation} problems are solvable in polynomial time -- in contrast to the consistency problems studied in the Section~\ref{subsec:hardness}.

Let us first prove that the candidate deletion evaluation problem is solvable in polynomial time, as it is the case for the corresponding consistency problem.

\begin{proposition}\label{prop:k-CD-eval}
\textsc{Candidate Deletion Single-Peaked Evaluation} can be solved in time $\mathcal{O}(n\cdot m^6)$.
\end{proposition}
\begin{proof}
Let $A$ be the given axis and $\mathcal{P}$ the profile for which we want to verify whether it is single-peaked with respect to $A$. Let $\mathcal{P}'$ be the profile obtained from $\mathcal{P}$ by adding two votes, $\vote{1}\,=A$ and $\vote{2}\,=\overline{A}$, where $\overline{A}$ denotes the reverse of axis $A$. By Lemma~\ref{lem:1..n,n..1}, $\mathcal{P}'$ is $k$-$X$ single-peaked consistent if and only if $\mathcal{P}$ is $k$-$X$ single-peaked with respect to $A$. By Lemma~\ref{lem:spc-subsets}, the same statement holds if $\mathcal{P}$ and $\mathcal{P}'$ are restricted to candidate subsets. Thus, applying the candidate deletion algorithm of Section~\ref{subsec:cd-poly-alg} to $\mathcal{P}'$ yields the same result as solving the evaluation problem for $\mathcal{P}$ and $A$.
\end{proof}

We now turn to evaluation problems where their corresponding consistency problems are \np-hard.

\begin{proposition}\label{prop:k-VD-eval}
\textsc{Voter Deletion Single-Peaked Evaluation} can be solved in time $\mathcal{O}(n\cdot m)$.
\end{proposition}
\begin{proof}
Whenever a vote is not single-peaked consistent with respect to $A$, we have to delete it.
If at most $k$ votes have to be deleted, we know that the profile is $k$-voter deletion single-peaked consistent with respect to $A$.
\end{proof}

\begin{proposition}\label{prop:k-AA-eval}
\textsc{Additional Axes Single-Peaked Evaluation} can be solved in time $\mathcal{O}(k\cdot n\cdot m)$.
\end{proposition}
\begin{proof}
Recall that for this evaluation problem all axes are given.
Hence it suffices to verify that every vote is single-peaked with respect to at least one of the given $k+1$ axes.
\end{proof}

\begin{theorem}\label{thm:k-LCD-eval}
\textsc{Local Candidate Deletion Single-Peaked Evaluation} can be solved in time $\mathcal{O}(n\cdot m^2\cdot \log m)$.
\end{theorem}
\begin{proof}
For every vote ${\vote{}}\in\mathcal{P}$ we have to find the minimum number of candidates that have to be deleted.
We do this by iterating over all candidates $p\in C$ and calculating a cardinality maximal $C'\subseteq C$ such that $\vote{}\![C']$ is single-peaked with respect to the given axis $A$ and $p$ is top-ranked in $\vote{}\![C']$.
For each $p$, let $C_1$ be the set of candidates containing $p$ and all candidates left of $p$ on $A$.
We have to find a (not necessarily contiguous) subsequence of $A[C_1]$ of maximum length that is 
increasing with respect to $\vote{}$ and contains $p$.
Let $C_1'$ be the set of candidates contained in this maximum increasing subsequence.
Then, let $C_2$ be the set of candidates containing $p$ and all candidates right of $p$.
We search for a subsequence of $A[C_2]$ of maximum length decreasing with respect to $\vote{}$ that contains $p$.
Let $C_2'$ be the set of candidates contained in this maximum decreasing subsequence.
In this way, we obtain a selection of candidates $C_1'\cup C_2'$ (of maximum cardinality) such that $\vote{}\![C_1'\cup C_2']$ is single-peaked and $p$ is top-ranked in $\vote{}\![C_1'\cup C_2']$.
We repeat this step for all $p\in C$ in order to find the candidate for which $\card{C_1'\cup C_2'}$ is maximal; let $d(\succ)$ denote number of required deletions for vote $\succ$, i.e., $d(\succ)=m-\card{C_1'\cup C_2'}$.
We return $\dlcd(\mathcal{P}) = \max_{\succ\in P} d(\succ)$.

Since finding a longest increasing subsequence in sequences of length $m$ can be done in time $\mathcal{O}(m\cdot\log m)$ \cite{schensted1961longest} and we have to do this $2m$ times per vote, we obtain the claimed runtime.
\end{proof}

\begin{theorem}\label{thm:k-GS/LS-eval}
\textsc{Global Swaps Single-Peaked Evaluation} as well as \textsc{Local Swaps Single-Peaked Evaluation} can be solved in time $\mathcal{O}(n\cdot m^3)$.
\end{theorem}
\begin{proof}
Both algorithms rely on the $\textsf{minswaps}(\vote{}, A)$ procedure, which computes the minimal number of swaps required to make vote $\vote{}$ single-peaked with respect to $A$.
Let us first describe how this procedure is used and later on give a precise description of $\textsf{minswaps}$.
To solve \textsc{Global Swaps Single-Peaked Evaluation} it suffices to execute for each $\vote{}$ in $\mathcal{P}$ the procedure $\textsf{minswaps}(\vote{}, A)$ and sum over all returned values.
If the sum does not exceed the limit $k$ we know that the profile is $k$-global swaps single-peaked consistent with respect to $A$.
For the \textsc{Local Swaps Single-Peaked Evaluation} the procedure is similar.
Here, we check for every $\vote{}$ in $\mathcal{P}$ whether $\textsf{minswaps}(\vote{}, A)\leq k$.

Let us now describe how $\textsf{minswaps}(\vote{}, A)$ can be implemented via dynamic programming.
In the following, let $\textit{sw}(c_i,\vote{}')$ be the minimum number of swaps required to move $c_i$ to the last position in $\vote{}'$, i.e., if if there are $k$ candidates ranked below $c_i$ then $\textit{sw}(c_i,\vote{}')=k$.
To simplify notation, we assume without loss of generality that $A$ is $c_1 > c_2 > \cdots > c_m$.
For $0\leq i<j\leq m+1$ let $s(i,j)$ be the minimal number of swaps required to transform a vote $\vote{}$ into a vote $\vote{}'$ such that $c_i \succ' c_{i-1} \succ' \cdots \succ' c_1$ as well as $c_j \succ' c_{j+1} \succ' \cdots \succ' c_m$, and all candidates in $\{c_{i+1},\ldots,c_{j-1}\}$ are ranked above all other candidates.
Observe that $s(i,i+1)$ is the number of swaps required to make $\succ$ single-peaked with respect to $A$ such that $c_i$ is the peak.
Thus, once we have values for $s(i,i+1)$ for all $1\leq i\leq m$, we can compute the number of swaps required to make $\succ$ single-peaked with respect to $A$ as $\min_{i\in\{0,\ldots,m\}}s(i,i+1)$.
Thus, $\textsf{minswaps}(\vote{}, A)$ returns $\min_{i\in\{0,\ldots,m\}}s(i,i+1)$.

The quantity $s(i,j)$ can be computed via dynamic programming.
Let $s(0,m+1)=0$~and
\begin{align*}
s(i,j) = \min( &s(i-1,j) + \textit{sw}(c_i,\vote{}\![\{c_i,\ldots,c_{j-1}\}]),\\
&s(i,j+1) + \textit{sw}(c_j,\vote{}\![\{c_{i+1},\ldots,c_{j}\}])),
\end{align*}
i.e., to compute $s(i,j)$ we can either start with $s(i-1,j)$ and move $c_i$ below $c_{i+1},\ldots,c_{j-1}$, or we start with $s(i,j+1)$ and move $c_j$ below $c_{i+1},\ldots,c_{j-1}$. In both cases $c_i \succ' c_{i-1} \succ' \cdots \succ' c_1$ as well as $c_j \succ' c_{j+1} \succ' \cdots \succ' c_m$, and the candidates in $\{c_{i+1},\ldots,c_{j-1}\}$ are ranked above all other candidates -- as we intend it to be.
We choose the option which requires~fewer~swaps.

Note that $s(i,j)$ has to be computed for $\mathcal{O}(m^2)$ values, which can be done in $\mathcal{O}(m^3)$ time.
Consequently, $\textsf{minswaps}$ can be computed in $\mathcal{O}(m^3)$ time.
We conclude that \textsc{Global Swaps Single-Peaked Evaluation} as well as \textsc{Local Swaps Single-Peaked Evaluation} can be solved in time $\mathcal{O}(n\cdot m^3)$.
\end{proof}

\begin{proposition}\label{prop:k-CP-eval}
\textsc{Candidate Partition Single-Peaked Evaluation} can be solved in time $\mathcal{O}(n\cdot m)$.
\end{proposition}
\begin{proof}
Let $C_1,\ldots, C_{k+1}$ be the given partition of $C$.
We can solve this problem in $\mathcal{O}(n\cdot m)$ time by verifying whether $\mathcal{P}[C_i]$ is single-peaked with respect to $A[C_i]$ for every $i\in\{1,\ldots,k+1\}$.
\end{proof}

Finally, one can show that the polynomial-time algorithms solving \textsc{Clones Single-Peaked Consistency} and \textsc{Width Single-Peaked Consistency} \cite{elk-fal-sli:c:clone-structures,DBLP:conf/ijcai/CornazGS13} are also applicable to the corresponding evaluation problems and maintain their time bounds; the argument for this is similar to the one of Proposition~\ref{prop:k-CD-eval}.

\section{Conclusions and Open Questions}
\label{sec:conclusions}

In this work, we have investigated notions of nearly single-peakedness.
We have introduced three new notions of nearly single-peakedness and have studied in addition six already established notions.
We have drawn a complete picture of the relations between these notions.
For five notions we have shown that deciding nearly single-peaked consistency is $\np$-complete and for $k$-candidate deletion we have presented a polynomial-time algorithm.
Furthermore, we have analyzed the complexity of the evaluation problem, i.e., the verification task, where the axis is given as additional input.
In contrast to consistency, all evaluation problems can be decided in polynomial time.
We refer the reader to Table~\ref{tab:results} for an overview.

An obvious direction for future work is to determine the complexity of \textsc{Candidate Partition Single-Peaked Consistency}.
Also, we want to remark that all notions of nearly single-peakedness presented in this work are not restricted to single-peakedness, but can also be applied to other domain restrictions (such as the single-crossing restriction).

\begin{table}
    \centering
    \begin{tabular}{lll}
        \toprule
        Notion & SP-Consistency & SP-Evaluation \\
        \midrule
        $k$-Voter Deletion & \np-c (Thm.~\ref{thm:k-voter-deletion-NPc}) & in \p\ (Prop.~\ref{prop:k-VD-eval}) \\
        $k$-Local Candidate Deletion & \np-c (Thm.~\ref{thm:k-LCD-NPc}) & in \p\ (Thm.~\ref{thm:k-LCD-eval}) \\
        $k$-Additional Axes & \np-c (Thm.~\ref{thm:k-AA-NPc}) & in \p\ (Prop.~\ref{prop:k-AA-eval}) \\
        $k$-Global Swaps & \np-c (Thm.~\ref{thm:k-GS-NPc}) & in \p\ (Thm.~\ref{thm:k-GS/LS-eval}) \\
        $k$-Local Swaps & \np-c (Thm.~\ref{thm:k-LS-NPc}) & in \p\ (Thm.~\ref{thm:k-GS/LS-eval}) \\
        $k$-Candidate Deletion & in \p\ (Thm.~\ref{thm:k-GD-poly}) & in \p\ (Prop.~\ref{prop:k-CD-eval}) \\
        $k$-Clones  & in \p~\cite{elk-fal-sli:c:clone-structures} & in \p~\cite{elk-fal-sli:c:clone-structures} \\
        $k$-Width  & in \p~\cite{DBLP:conf/ijcai/CornazGS13} & in \p~\cite{DBLP:conf/ijcai/CornazGS13} \\
        $k$-Candidate Partition  & open & in \p\ (Prop.~\ref{prop:k-CP-eval}) \\
        \bottomrule
    \end{tabular}
    \caption{Complexity results for different notions of nearly single-peakedness.}
    \label{tab:results}
\end{table}

$\np$-completeness, as we have obtained for several consistency problems, does not rule out the possibility of algorithms that perform well in practice.
One approach is to search for fixed-parameter algorithms, i.e., an algorithm with runtime $f(k)\cdot \mathit{poly}(n)$ for some computable function $f$ depending only on parameter $k$.
A first fixed-parameter algorithm for \textsc{Voter Deletion Single-Peaked Consistency} is mentioned by \citeA{bre-che-woe-profiles-nearby}.
Another approach is the development of approximation algorithms since nearly single-peaked consistency can also be seen as an optimization problem.
A detailed treatment of both fpt- and approximation-results for Candidate Deletion and Voter Deletion was presented recently for several domain restrictions including single-peakedness \cite{aaai/ElkindL14-approx}.
The design of fixed-parameter and approximation algorithms for the remaining notions of nearly single-peakedness deserves further attention.

Another interesting direction for future work is extending our models to manipulative behavior, such as manipulation, control, and bribery.
That is, assuming we have a nearly single-peaked electorate according to one of our notions, how computationally expensive is a manipulative action under a certain voting rule?
The analysis of manipulation and control in such elections has already been started for some distance measures.
In a first step, manipulation and control was introduced in the context of nearly single-peaked elections under several voting rules, pinpointing that under some voting rules even the presence of only one maverick can raise the complexity of manipulative actions from $\p$ to $\np$-completeness~\cite{fal-hem-hem:j:nearly-single-peaked}. In a second step, dichotomy results for manipulation of $k$-approval and $k$-veto under nearly single-peaked elections were achieved identifying the exact borders of tractability~\cite{adt/ErdelyiLP15-manipulation}.
Still, the impact of nearly single-peakedness on manipulative behavior is far from being fully understood.

Finally, there might be further useful and natural distance measures regarding single-peakedness to be found.
In particular, for practical purposes it may be useful to combine distance measures. For example, for a real-world preference data set two axes may be sufficient if in addition a small number of swaps is allowed. Computing distances of this sort would be require to define how to weigh measures against each other, e.g., how does an additional axis compare to 10 swaps? In addition, practical considerations have to be made to decide which distance measures are useful to combine. The aforementioned combination of Global Swaps and Additional Axes may be useful since it allows to correct minor disturbances in the preferences (``noise'') but also take a segregation of voters into account (e.g., there are two issues that are deemed important; every voter gives preference to one and orders the candidates accordingly).
The hardness results obtained in this paper do not necessarily transfer to consistency problems for such ``mixed'' distance measures.
These thoughts do not only apply to the single-peaked restriction but also to other domain restrictions such as the single-crossing or more-dimensional analogues of single-peakedness.

\section*{Acknowledgments}
The third author, Andreas Pfandler, is affiliated both with the University of Siegen and TU Wien. 
This work was done in part while the second and the third author were visiting the University of Siegen and while the first author was visiting TU Wien, the University of Rochester, and NICTA, Sydney. This work was supported in part by the Austrian Science Fund (FWF) under grant P25518-N23 and Y698, by DFG grants ER 738/1-1 and ER 738/2-1, by COST Action IC1205, and by the European Research Council (ERC) under grant number 639945 (ACCORD).
We thank Dominik Peters for pointing out a mistake in an earlier version of this paper.
We would also like to thank the anonymous JAIR referees for their very helpful comments and suggestions as well as the reviewers of earlier versions of this paper \cite{erd-lac-pfa:c:single-peaked,erd-lac-pfa:c:single-peaked-aaai}.

\appendix

\section{Correctness of the \emph{Place} Procedure}

\lemplacecorrect*

\begin{proof}
Let us assume towards a contradiction that the iterative application of $\textsf{place}$ yields an axis $A$, but $(C,\mathcal{P})$ is not single-peaked with respect to $A$.
By Lemma~\ref{lem:not-spc-crit}, there exists a vote ${\vote{}}$ in $\mathcal{P}$ and $c_i, c_j, c_k\in C$ such that $c_i>c_j>c_k$ on $A$, $c_i\succ c_j$ and $c_k\succ c_j$.
Let us consider the step in the algorithm at which the last element of $\{c_i, c_j, c_k\}$ is placed.
Let $A'$ be the incomplete axis at this point with $\textsf{boundary}(A')=(b_1',b_1,b_2,b_2')$.
We assume that the election is single-peaked with respect to $A'$, i.e., we consider the earliest step at which $\textsf{place}$ makes a mistake.
Before we start with the proof, we prove the following useful claim.

\medskip
\noindent \textbf{Claim A.}
Let $A'$ be an incomplete axis with $\textsf{boundary}(A')=(b_1',b_1,b_2,b_2')$ and $\succ$ is single-peaked with respect to $A'$.
If for some $c,c'$ it holds that $c\succ c'$ and $c>c'\geq b_1$ (i.e., $c'$ might be the same candidate as $b_1$), then $b_1'\succ b_1$.
Analogously, if $c\succ c'$ and $b_2\geq c'> c$, then $b_2'\succ b_2$.
\smallskip
\vspace*{-0.5em}
\begin{proof}
We prove the first statement; the second one is analogous.
Note that if $c=b_1'$ then $c'=b_1$ and the statement is immediate.
So assume that $c\neq b_1$.
Since $c> c'\geq b_1' > b_1$, $c\succ c'$ and $b_1\succ b_1'$ contradict single-peakedness.
\end{proof}

\noindent To disprove our assumption that $(C,\mathcal{P})$ is not single-peaked with respect to $A$, let us distinguish the following cases.
\begin{itemize}
\item $c_i$ has been placed last:
When placing $c_i$, it holds that $b_2\geq c_j>c_k$ on $A'$. Since $c_k\succ c_j$, Claim A implies that $b_2'\succ b_2$.
However if $b_2'\succ b_2$ and $c_i\succ c_j \succeq b_2$, then by Case 4 $\textsf{place}$ returns $\texttt{inconsistent}$, a contradiction to our assumption that $\textsf{place}$ returns an axis. The case that $c_k$ is placed last is symmetric.
\item $c_j$ has been placed last: It is not possible that $b_1\succ c_j$ and $b_2\succ c_j$ since $c_j$ would have been placed at an earlier point (cf.~Case 3a). Thus either $c_j\succ b_1$ or $c_j\succ b_2$. If $c_j\succ b_1$, then by Claim A $b_1'\succ b_1$; if $c_j\succ b_2$, then by Claim A $b_2'\succ b_2$. Hence by Case 4 $\textsf{place}$ returns $\texttt{inconsistent}$.
\item $\{c_i, c_j\}$ have been placed last: Let us first prove that $b_2\succ c_j$. If $b_2=c_k$, then $b_2=c_k\succ c_j$. If $b_2\neq c_k$ and $c_j\succ b_2$, then $c_k\succ b_2$ and by Claim A $b_2'\succ b_2$. Thus by Case~4 $\textsf{place}$ would return $\texttt{inconsistent}$, a contradiction.
We have shown that $b_2\succ c_j$. 
But if $b_2\succ c_j$ and $c_i\succ c_j$ (by assumption), then by Case~2c $c_j$ has to be placed next to $b_1$ and $c_i$ next to $b_2$, and hence $c_j > c_i > c_k$ on $A$. This contradicts our assumption that $c_i>c_j> c_k$ on $A$. The case that $\{c_j, c_k\}$ are placed last is symmetric.
\end{itemize}
We conclude that $(C,\mathcal{P})$ is single-peaked with respect to $A$.

Now we show that if $\textsf{place}$ returns $\texttt{inconsistent}$, then $(C,\mathcal{P})$ is not single-peaked.
Let $A'$ be an incomplete axis, let $C'$ be the candidates already placed on $A'$ and let $X$ the set of candidates to be placed next, i.e., $X=L(\mathcal{P},C\setminus C')$.
Assume that $\textsf{place}(A',X)$ returns $\texttt{inconsistent}$.
We consider all cases where $\textsf{place}$ returns $\texttt{inconsistent}$ and show that $(C,\mathcal{P})$ is not single-peaked consistent.
Since we have already shown that the iterative application of $\textsf{place}$ yields a single-peaked axis, we can assume that $\mathcal{P}[C']$ is single-peaked with respect to $A'$.
\begin{itemize}
\item Case 1: It follows from Lemma~\ref{lem:spc-subsets} and \ref{lem:three-last-ranked} that $(C,\mathcal{P})$ is not single-peaked consistent.
\item Case 4: Let $\succ_1$ be the vote under consideration with $b_1'\succ_1 b_1$ and $x\succ_1 b_1$. Let us consider the step when $b_1$ was placed; the incomplete axis at this point was of the form $\dots>b_1'>\star>\cdots$. Since $b_1'\succ_i b_1$ and $b_1$ was placed next to $b_1'$, we know that Case 2d occurred. (If Case 3 had been applicable then $b_1'\succ_i b_1$ would imply that $b_1$ is placed on the right-hand side and not next to $b_1'$.) Let $y$ be the second candidate that was placed at this step. Since the profile is single-peaked with respect to the incomplete axis $A'$, we know that $b_1'\succ_1 b_1\succ_1 y$. 
Since $b_1$ was placed at this step, there has to be a vote $\succ_2$ with $y\succ_2 b_1$. Furthermore, it has to hold that either $x\succ_2 y$ or $x\succ_2 b_1$ since $x$ was not placed at this step. Finally, $b_1\succ_2 b_1'$ has to hold, because otherwise the profile would not be single-peaked with respect to $A'$. To sum up, we have $b_1'\succ_1 b_1\succ y$ and $x\succ_1 b_1$ as well as $y\succ_2 b_1\succ_2 b_1'$ and $x\succ_2 b_1$. It is straight-forward to verify that this subprofile is not single-peaked; indeed, this is one of the forbidden subprofiles occurring in the single-peakedness characterization by \citeA{bal-hae:j:characterization-single-peaked}. Hence by Lemma~\ref{lem:spc-subsets} the full profile is not single-peaked. 

\item The procedure may also return $\texttt{inconsistent}$ if a candidate $x_1$ has to be placed both left and right. This occurs if (i) Case 2c occurs once as stated and once with $x_1$ and $x_2$ interchanged, (ii) Case 2d occurs once as stated and once with $x_1$ and $x_2$ interchanged, (iii) both Case 2c and 2d occur, and (iv) both Case 2c and 2d occur with $x_1$ and $x_2$ interchanged. We first handle (i); the argument for (ii) is analogous. Let $L(\mathcal{P},C') = \{x_1,x_2\}$. Assume that there exists a vote $\succ_1$ with $b_2\succ_1 x_1$ and $x_2\succ_1 x_1$ (Case 2c). Further there exists a vote $\succ_2$ with $b_2\succ_2 x_2$ and $x_1\succ_2 x_2$. Since $b_2$ has been placed at an earlier step, there exists a vote $\succ_3$ where $x_1\succ_3 b_2$ and $x_2\succ_3 b_2$. This proves that the profile is not single-peaked via Lemma~\ref{lem:three-last-ranked}.

Let us now consider (iii); the argument for (iv) is analogous. There exists a vote $\succ_1$ with $b_2\succ_1 x_1$ and $x_2\succ_1 x_1$ (Case 2c). Since $b_1$ and $b_2$ have been placed before $x_1$ and $x_2$, either $b_1$ or $b_2$ has to be ranked below both $x_1$ and $x_2$. Thus it has to hold that $x_2\succ_1 x_1\succ_1 b_1$. Also, there exists a vote $\succ_2$ exists with $b_1\succ_2 x_1$ and $x_2\succ_2 x_1$ (Case 2d) and by the same argument as before $x_2\succ_2 x_1\succ_2 b_2$. It is straight-forward to show that this subprofile is not single-peaked; indeed, it is equivalent to the one we encountered when we dealt with Case 4. Hence the full profile is not single-peaked.

\item Let us consider contradictions that can arise in Case 3. Recall from the algorithm description that a contradiction can only arise if $x$ is not the last candidate to be placed. Let $y\neq x$ be some candidate not yet placed. It has to hold that $y\succ x$ in all votes.
Let $\succ_1$ be a vote with $b_2\succ_1 x$. Since both $b_1$ and $b_2$ have been placed before $x$, we infer that $b_2\succ_1 x\succ_1 b_1$. Further let $\succ_2$ be a vote with $b_1\succ_2 x$ and hence $b_1\succ_2 x\succ_2 b_2$. Considering the votes $\succ_1,\succ_2$ and candidates $\{x,y,b_1,b_2\}$, we encounter the same subprofile as in the previous step.
\end{itemize}
This concludes the correctness proof of the $\textsf{place}$ procedure.
\end{proof}

\bibliographystyle{theapa}
\bibliography{single_peak}

\end{document}